\documentclass[12pt]{amsart}
\usepackage[margin=1in,letterpaper]{geometry} 

\usepackage{times}

\usepackage{amssymb,amsfonts,amsmath,bbm,mathrsfs,stmaryrd}

\usepackage{xcolor}
\usepackage[colorlinks,
             linkcolor=black!50!red,
             citecolor=blue,
             pdftitle={Homological perturbation theory for nonperturbative integrals},
             pdfauthor={Theo Johnson-Freyd},
             pdfsubject={integrals},
             pdfkeywords={integrals,homological perturbation},
             pdfproducer={pdfLaTeX},
             pdfpagemode=None,
             bookmarksopen=true
             bookmarksnumbered=true]{hyperref}

\newcommand\arXiv[1]{\href{http://arxiv.org/abs/#1}{\nolinkurl{arXiv:#1}}}
\newcommand\MRnumber[1]{\href{http://www.ams.org/mathscinet-getitem?mr=#1}{\nolinkurl{MR#1}}}
\newcommand\DOI[1]{\href{http://dx.doi.org/#1}{\nolinkurl{DOI:#1}}}
\newcommand\MAILTO[1]{\href{mailto:#1}{\nolinkurl{#1}}}

\newcommand\longto{\longrightarrow}
\newcommand\<\langle
\renewcommand\>\rangle
\newcommand\sminus{\smallsetminus}
\renewcommand\AA{{\mathbb A}}
\newcommand\Oo{{\mathcal O}}
\newcommand\CC{{\mathbb C}}
\newcommand\Ccc{{\mathscr C}}
\newcommand\Ll{{\mathcal L}}

\newcommand\PP{{\mathbb P}}
\newcommand\RR{{\mathbb R}}
\newcommand\ZZ{\mathbb Z}
\newcommand\T{{\mathrm T}}
\renewcommand{\d}{{\mathrm d}}

\newcommand{\id}{\mathrm{id}}

\newcommand\dVol{\mathrm{dVol}}
\DeclareMathOperator\homology{H}
\renewcommand\H\homology
\DeclareMathOperator\Sym{Sym}

\newcommand\cl{{\mathrm{cl}}}

\newcommand\diag{{\mathrm{diag}}}
\newcommand\mix{{\mathrm{mix}}}

\newcommand\vv{\vec v}
\newcommand\BV{{\mathrm{BV}}}
\newcommand\MV{{\mathrm{MV}}}

\DeclareMathOperator\divergence{div}
\renewcommand\div\divergence
\renewcommand\i{\sqrt{-1}}
\renewcommand\[{\llbracket}
\renewcommand\]{\rrbracket}
\DeclareMathOperator\gr{gr}
\newcommand\Ggg{\mathscr G}

\newcommand{\define}[1]{{\em #1}}

\usepackage{tikz}

\usetikzlibrary{arrows,decorations.pathmorphing,decorations.pathreplacing,decorations.markings,shapes.geometric,through,fit,shapes.symbols}

\tikzset{
    dot/.style={circle,draw,fill,inner sep=1pt},
    fdot/.style={circle,draw,fill,inner sep=1.5pt},
    Odot/.style={circle,draw,inner sep=1pt},
    Stardot/.style={star, star point height=2pt, inner sep=.5pt, draw, fill},
    twice/.style={double, double distance=1pt},
    Feynman/.style={color=green!50!black,text=black},
    derivative/.style={draw,dotted,thick,color=purple!50!blue,text=black},
    cloudsum/.style={cloud,draw,blue,text=black,inner sep=1pt},
    Feynmannode/.style={color=blue,text=black,draw},
    onearrow/.style={postaction={decorate}, decoration={markings,mark=at position .6 with {\arrow[draw,line width=1pt]{>}}}},
    inversearrow/.style={postaction={decorate}, decoration={markings,mark=at position .45 with {\arrow[draw,line width=1pt]{<}}}},
    twoarrows/.style={draw, postaction={decorate}, decoration={markings,mark=at position .35 with {\arrow[draw,line width=1pt]{>}},mark=at position .75 with {\arrow[draw,line width=1pt]{>}}}},
    twoarrowsempty/.style={postaction={decorate}, decoration={markings,mark=at position .3 with {\arrow[draw,line width=1pt]{>}},mark=at position .7 with {\arrow[draw,line width=1pt]{>}}}},
    inversetwoarrows/.style={draw, postaction={decorate}, decoration={markings,mark=at position .35 with {\arrow[draw,line width=1pt]{<}},mark=at position .7 with {\arrow[draw,line width=1pt]{<}}}},
    squiggly/.style={draw, decorate,decoration={snake,amplitude=.2mm,segment length=2mm}},
    inversesquiggly/.style={draw, decorate,decoration={snake,amplitude=.2mm,segment length=2mm},postaction={decorate,decoration={markings,mark=at position .45 with {\arrow[draw,line width=1pt]{<}}}}},
    Xobject/.style={color=red!75!black,squiggly},
    Aobject/.style={color=blue!50!red,ultra thick},
    Cobject/.style={color=blue},
    degreeshift/.style={onearrow,dashed},
}

\title{Homological perturbation theory for nonperturbative integrals}
\author{Theo Johnson-Freyd}

\begin{document}
\maketitle

\begin{abstract}
  We use the homological perturbation lemma to produce explicit formulas computing the class in the twisted de Rham complex represented by an arbitrary polynomial.  This is a non-asymptotic version of the method of Feynman diagrams.  In particular, we explain that phenomena usually thought of as particular to asymptotic integrals in fact also occur exactly: integrals of the type appearing in quantum field theory can be reduced in a totally algebraic fashion to integrals over an Euler--Lagrange locus, provided this locus is understood in the scheme-theoretic sense, so that imaginary critical points and multiplicities of degenerate critical points contribute.
\end{abstract}

\section{Introduction}

The method of Feynman diagrams computes, in a totally algebraic fashion, the asymptotics of compactly supported oscillating integrals: the input consists of the power-series expansions of the ``action'' and ``observable'' functions in question; the Feynman diagrams encode rational functions in these Taylor coefficients; only at one step is any transcendental input required, and it is only to know the volume of a Gaussian distribution (some power of $\pi$).  

This paper presents analogous exact formulas for non-asymptotic integrals, in an algebrogeometric setting.  We will use the homological perturbation lemma to reduce any ``oscillating'' integral to an integral over the scheme-theoretic critical locus of the ``action'' function.  This reduction step consists of  explicit rational functions of the coefficients of the action, and results in a linear combination of finitely many integrals that seem to be irreducibly transcendental.  We will allow critical points with high-order degeneracy, although we will not work in the most general setting --- we focus instead on the simplest case of bosonic integrals in finitely many variables, and the reader is invited to adapt our techniques to whatever problem is at hand.

\subsection{Outline of the paper}

In Section~\ref{section.BVmotivation}, we give some motivation for the Batalin--Vilkovisky approach to ``oscillating'' integrals of the shape $\int_{X} f\,e^{s}\,\dVol$, and overview the construction of the (quantum) BV chain complex.  Most of the section is devoted to Example~\ref{eg.Feynman}, in which we review from~\cite{GJF} the derivation of Feynman diagrams from the BV complex.  In Example~\ref{eg.main} we introduce the main topic of this paper, namely integrals in which the functions $f$ and $s$ are complex polynomials, and we briefly discuss contours of integration.

The meat of the paper is in Section~\ref{section.polynomials}.  For the impatient readers who skipped Section~\ref{section.BVmotivation}, we begin by precisely defining the classical and quantum Batalin--Vilkovisky complexes that we will be concerned with.  The classical BV complex for a function $s\in \Oo(X)$ is the Koszul resolution of the critical locus of $s$, and the quantum BV complex is a version of the twisted de Rham complex.  The basic question is to have explicit control over the quantum BV complex: the algebraic part of ``integration'' is the problem of computing the homology class represented by a given $f\in \Oo(X)$.  In Theorem~\ref{thm.main}, we give explicit formulas identifying the quantum BV homology with the classical critical locus, thereby providing an algebraic way to ``integrate out the higher modes,'' analogous to the method of Feynman diagrams used in asymptotic integration.  Our basic tool is the  Homological Perturbation Lemma~\ref{hpl}, which is a well-known formula-full version of spectral sequences.

We conclude the paper with Section~\ref{section4}, which discusses at a non-rigorous level to what extent our techniques can be applied to the infinite-dimensional integrals appearing in quantum field theory, and we include some brief comments on the Volume Conjecture in Chern--Simons Theory.

\subsection{Acknowledgments} 

I have had valuable conversations about this material with Tom Goodwillie, Owen Gwilliam, Nicolai Reshetikhin, and Shamil Shakirov.
  This work is supported by the NSF grants DMS-0901431 and DMS-1304054.

\section{Encoding oscillating integrals as the Batalin--Vilkovisky chain complexes} \label{section.BVmotivation}

The close relationship between homological algebra and integration theory has been known since at least the time of de Rham (a good history is available in~\cite{MR1721123}).  In this section, we describe in general terms the shape of integrals appearing in quantum field theory.  The values of such integrals are controlled by a \define{twisted de Rham complex} (for various reasons, we will use instead the term ``Batalin--Vilkovisky complex''), which is the universal recipient for an ``integral'' satisfying Stokes' formula.  Specializing to the formal asymptotic case gives a complex for which a straightforward analysis results in the method of Feynman diagrams; details are in Example~\ref{eg.Feynman}.  In Section~\ref{section.polynomials} we will apply a similar analysis to the polynomial case, which we will briefly introduce in Example~\ref{eg.main}, where we also comment on the space of contours.

\subsection{``Oscillating'' integrals}

It is a basic tenet of quantum field theory that the predicted values of physical measurements should occur as the values of definite integrals (see e.g.~\cite{MR2478121}).   Specifically, in quantum field theory there is a space $X$ of ``fields'' (which is usually an infinite-dimensional derived stack) equipped with a ``volume form'' $\dVol$ (often $\dVol$ is the unique-up-to-scale ``volume form'' compatible with some symmetry).  The ``physics'' is controlled by an \define{action function} $s\in \Oo(X)$, where  $\Oo(X)$ denotes some distinguished algebra of ($\CC$-valued, say) functions on $X$.  The data of a measurement is encoded in an \define{observable} $f\in \Oo(X)$, and the predicted expectation value of the measurement is the ratio of definite integrals $\<f\>_{s} = I_{s}(f) / I_{s}(1)$, where $I_{s}(f) = \int_{X} f\,e^{s}\,\dVol$.  One special case is when $X$ is a real manifold and $s$ is pure-imaginary.  Then the integrand is oscillatory, and we will  refer to any integral of the shape $I_{s}(f) = \int_{X} f\,e^{s}\,\dVol$ as an \define{oscillating integral} by analogy.

The most physically interesting question is the ``inverse problem'': one measures the values of $I_{s}(-)$ for various inputs $f$, and tries to extract from this data information about the action $s$.  To get off the ground, however, one must begin by understanding how to compute $I_{s}(f)$ given the data of $s$ and $f$.  Or perhaps one should say ``one must begin by understanding how to \emph{define} $I_{s}(f)$,'' as the spaces appearing in quantum field theory tend not to be of the type that support analytic definitions of integration.  Whether the problem is one of computation or definition, the best situation is when the functional $I_{s}$ depends entirely algebraically on $s$: algebraic definitions can more easily be transported to infinite dimensions, and are more readily computable.

\subsection{Approximate Definition (BV Complex)}

Fix $(X,\dVol)$ and $s$.  Any reasonable theory of integration should satisfy a \define{Stokes formula}, computing the value of $I_{s}(-)$ on ``total derivatives'' in terms of a ``boundary term.''  In many situations, these boundary terms can be made to vanish identically.  If so, $I_{s} : \Oo(X) \to \CC$ factors through the quotient vector space $\Oo(X) / \{\text{total derivatives}\}$.  Often this quotient is finite-dimensional, and so our approach to understanding $I_{s}$ will be to understand algebraically the projection $\Oo(X) \to \Oo(X) / \{\text{total derivatives}\}$ with respect to some distinguished bases, so that the only non-algebraic part of the functional $I_{s}$ consists of finitely many data describing the map $\Oo(X) / \{\text{total derivatives}\} \to \CC$.

The \define{quantum Batalin--Vilkovisky (or BV) complex} is a chain complex that resolves the quotient $\Oo(X) / \{$total derivatives$\}$.  It is constructed as follows.  Let $V$ be some vector space consisting of vector fields on $X$ that are divergence-free for the volume form $\dVol$.  Then a \define{total derivative} is a function of the form $\vv(g\,e^{s}) = \bigl(\vv(g) + g\,\vv(s)\bigr)e^{s}$ for $g\in \Oo(X)$ and $\vv\in V$.  Thus the vector space $\{\text{total derivatives}\}$ is the image of the map $\partial_{s} : V\otimes \Oo(X) \to \Oo(X)$ given by $(\vv,g) \mapsto \vv(g) + g\,\vv(s)$, and so we can compute $\Oo(X) / \{$total derivatives$\}$ as the degree-$0$ homology of a two-term chain complex
$$ \frac{\Oo(X) }{ \{\text{total derivatives}\}} = \H_{0}\biggl( V\otimes \Oo(X) \overset{\partial_{s}}\longto \Oo(X) \biggr), $$
where $V\otimes \Oo(X)$ is in homological degree $1$ and $\Oo(X)$ is in degree $0$.  This two-term complex has extra homology in degree $1$, and so we construct the complete BV complex in the usual ``Koszul'' way by taking an exterior power:
\begin{gather*}
  \BV_{\bullet}(X,\dVol,s) = V^{\wedge \bullet} \otimes \Oo(X), \\
  \partial_{\BV}(\vv_{0}\wedge \cdots \wedge \vv_{k-1} \otimes g) = \sum_{i=0}^{k-1} (-1)^{i}\, \vv_{0} \wedge \cdots  \widehat{\vv_{i}} \cdots \wedge \vv_{k-1} \otimes (\vv_{i}(g) + g\,\vv_{i}(s))
\end{gather*}
The ``hat'' denotes removing the $i$th term in the wedge product.

This definition is only approximate, because details like the choice of $V$ might depend on one's application.  Rather than trying to give a completely precise universal definition, we will illustrate the notion of Batalin--Vilkovisky complex with Examples~\ref{eg.smooth}, \ref{eg.Feynman}, and~\ref{eg.main}.

\subsection{Example (smooth manifolds)}\label{eg.smooth}

Suppose that $X$ is an $n$-dimensional compact oriented smooth manifold equipped with a nowhere-vanishing top-form $\dVol \in \Omega^{n}(X)$, where $\Omega^{\bullet}(X)$ denotes the smooth de Rham complex.  Consider the graded vector space $\MV_{\bullet}(X) = \Gamma\bigl(\T^{\wedge \bullet}X\bigr)$ of antisymmetric multivector fields, constructed from the tangent bundle in analogy to the construction of $\Omega^{\bullet}(X) = \Gamma\bigl((\T^{*})^{\wedge \bullet}X\bigr)$ from the cotangent bundle.  The choice of $\dVol$ determines an isomorphism $\MV_{\bullet}(X) \cong \Omega^{n-\bullet}(X)$ given by contraction with $\dVol$.  The  differential on $\MV_{\bullet}$ given by pulling the de Rham differential $\d$ across this isomorphism is the \define{divergence} operator on multivector fields; on vector fields, it is defined by $\div(\vv)\, \dVol = \Ll_{\vv}(\dVol)$, where $\Ll_{\vv}$ denotes the Lie derivative in the $\vv$ direction.  Given a function $s\in \Ccc^{\infty}(X)$, we get an exact one-form $\d s$ and a corresponding map $\iota_{\d s} : \MV_{\bullet}(X) \to \MV_{\bullet-1}(X)$ given by contraction with $\d s$.
The BV complex is:
$$ \BV_{\bullet}(X,\dVol,s) = \MV_{\bullet}(X), \quad \partial_{\BV} = \div + \iota_{\d s} $$
Note that the data of $\div$ and $\iota_{\d s}$ are invariant under rescaling $\dVol$ or shifting $s$ by any locally constant function, so that the \define{BV differential} $\div + \iota_{\d s}$ depends only on the projective data of $e^{s}\,\dVol$; on the other hand, one can recover $e^{s}\,\dVol$ up to locally-constant rescaling from $\div + \iota_{\d s}$.

If $X$ is not compact, then the most natural functions $f$ to integrate are the compactly-supported ones, and in order to assure that there are no boundary terms we should request compact-support in higher degrees as well.  So we take $\BV_{\bullet}$ to consist of the compactly-supported multivector fields, with the same differential.  If $X$ is not oriented, $\dVol$ is not a section of $\Omega^{n}(X)$ but rather of the density line bundle, and so the interpretation in terms of the de Rham complex must be corrected by an orientation bundle; we can still define $\BV_{\bullet}$ in terms of multivector fields and their divergences.

This example is elaborated upon in~\cite{WittenAntibracket}.

\subsection{Remark (twisted de Rham complex)} \label{remark.twisted}

When $X$ is an $n$-dimensional compact oriented manifold, Example~\ref{eg.smooth} describes a Batalin--Vilkovisky complex $\bigl(\MV_{\bullet}(X),\partial_{\BV}\bigr)$ that is closely related to the de Rham complex $\bigl(\Omega^{\bullet}(X),\d\bigr)$, where as always $\d$ denotes the de Rham differential.  In fact, by contracting not with $\dVol$ but with $e^{s}\,\dVol$, one can construct an isomorphism of chain complexes $\bigl(\MV_{\bullet}(X),\partial_{\BV}\bigr) \cong \bigl(\Omega^{n-\bullet}(X),\d\bigr)$.  The isomorphism ``contract with $\dVol$'' used in Example~\ref{eg.smooth} instead relates $\partial_{\BV}$ to the differential $\d +  \d s\wedge$ on $\Omega^{n-\bullet}(X)$, where $ \d s\wedge$ is the operation of multiplication by the exact one-form $\d s$.  

The cochain complex $\bigl( \Omega^{\bullet}(X), \d +  \d s\wedge\bigr)$ is called the \define{twisted de Rham complex} for $s$.  Of course, the differential $\d +  \d s\wedge$ is the result of conjugating $\d$ by the operation of multiplication by $e^{s}$, and so  for smooth or analytic manifolds the twisted and untwisted de Rham complexes are isomorphic.
For most of this paper we will be interested in the polynomial situation when $e^{s} \not\in \Oo(X)$.  Then the twisted and untwisted de Rham complexes are not in general isomorphic.  Up to a grading shift, our Batalin--Vilkovisky complex will remain isomorphic to the twisted de Rham complex.

Nevertheless, we prefer not to use the language of twisted de Rham complexes.  There are two main reasons for this:
\begin{enumerate}
  \item  In the Batalin--Vilkovisky complex, the homology group of most interest is in degree $0$, whereas in the de Rham complex the most interesting homology group is in ``top'' degree.  For finite-dimensional spaces, this is a mild \ae sthetic difference, but it becomes important when trying to generalize to the infinite-dimensional spaces appearing in quantum field theory: the natural infinite-dimensional de Rham complex includes functions, one-forms, two-forms, etc., and has no ``top,'' whereas the natural infinite-dimensional Batalin--Vilkovisky complex has no ``bottom.''  
  
%  A related issue is that when $X$ fails to be both compact and oriented, the most natural definitions of the twisted de Rham complex and the Batalin--Vilkovisky complex diverge.
  \item  Although it won't play a major r\^ole in this paper, Batalin--Vilkovisky complexes have important algebraic structures that are obscured by thinking of them in terms of twisted de Rham complexes.  Both $\MV_{\bullet}(X)$ and $\Omega^{\bullet}(X)$ are graded-commutative rings under wedge multiplication.  The de Rham differential $\d$ is a derivation of the algebra structure on $\Omega^{\bullet}(X)$, making $\bigl(\Omega^{\bullet}(X),\d\bigr)$ into a dga; the twisted differential $\d + \d s \wedge$ is not a derivation, but is a first-order differential operator.  On the other hand, on $\MV_{\bullet}(X)$ the differential $\iota_{\d s}$ is a derivation, and for the wedge multiplication the ``BV'' differential $\div$ is a {second-order} differential operator.  (Recall that an endomorphism of a graded-commutative algebra is a \define{$k$th-order differential operator} if its graded-commutator with multiplication by any element of the algebra is a $(k-1)$th-order differential operator.)  Moreover, the failure of $\div$ to be a derivation for $\wedge$ is measured exactly by the \define{Schouten--Nijenhuis bracket} on $\MV_{\bullet}(X)$, which is an important piece of classical geometry.
  
  Often for more general $X$, the graded-commutative algebra $\MV_{\bullet}(X)$ can be defined and  equipped with a Schouten--Nijenhuis bracket.  A \define{BV projective volume form} on $X$ is  a differential on $\MV_{\bullet}(X)$ whose failure to be a derivation is measured by the Schouten--Nijenhuis bracket.  Our techniques for writing down explicit formulas should extend well to many  settings in which a BV projective volume form is given.
  
  What will be important in this paper is to relate the BV complex $\bigl(\MV_{\bullet}(X),\div + \iota_{\d s}\bigr)$ with the \define{derived critical locus} $\bigl(\MV_{\bullet}(X),\iota_{\d s}\bigr)$, which is a dga whose degree-$0$ homology is the algebra of functions on the critical locus of $s$.  In this way, the algebraic structure on the Batalin--Vilkovisky complex will be directly important.
\end{enumerate}
Further discussion of the ``BV philosophy'' is available in many places, including \cite{Stasheff:fk,Fiorenza04,ABF,MR2778558,costellogwilliam}.

\subsection{Extended Example (Feynman diagrams)} \label{eg.Feynman}

The Batalin--Vilkovisky formalism packages neatly the method of Feynman diagrams, as we now explain.  This section is based on~\cite{GJF}.

Suppose that $f$ is a compactly supported complex-valued smooth function on a manifold $X$ equipped with volume form $\dVol$, and that $s$ is smooth and real-valued.  We denote the critical locus of $s$ by $\{\d s = 0\}$.  Suppose that $\{\d s = 0\}$ is \define{clean}, in the sense that it is an embedded submanifold of $X$ and that the Hessian of $s$ is nondegenerate in directions transverse to $\{\d s = 0\}$.  For simplicity, suppose furthermore that $s$ takes constant-value $0$ on $\{\d s = 0\}$.  Then as $\hbar\to 0$ along positive real numbers, the values of the integrals $\int_{X} f\exp\bigl(\i s/\hbar\bigr)\,\dVol$ have an asymptotic expansion as a power series in $\hbar$.  Moreover, this expansion depends only on the value of $f$ near $\{\d s = 0\}$.  For details on these estimates, see~\cite{MR2952218}.

Thus, if we are interested only in the $\hbar\to 0$ asymptotics of $\int_{X} f\exp\bigl(\i s/\hbar\bigr)\,\dVol$, we can choose a tubular neighborhood $X_{\approx\{\d s = 0\}}$ of $\{\d s = 0\}$ and multiply $f$ by a bump function supported in this neighborhood.  We can then choose a contraction making $X_{\approx\{\d s = 0\}}$ into a fiber bundle over $\{\d s = 0\}$.  Pushing the density $f\exp\bigl(\i s/\hbar\bigr)\,\dVol$ forward along $X_{\approx\{\d s = 0\}} \to \{\d s = 0\}$ produces a (smooth!) density $\int_{\mathrm{fiber}}f\exp\bigl(\i s/\hbar\bigr)\,\dVol$ on $\{\d s = 0\}$, which we can then integrate over $\{\d s = 0\}$ to produce $\int_{X} f\exp\bigl(\i s/\hbar\bigr)\,\dVol$.  As $\hbar\to 0$, the density $\int_{\mathrm{fiber}}f\exp\bigl(\i s/\hbar\bigr)\,\dVol$ makes sense as a $\CC\[\hbar\]$-valued measure on $\{\d s = 0\}$.  In physics, one would consider this pushed-forward measure as an ``effective action'' achieved by ``integrating out the high-energy modes.''

Let $s^{(2)}$ denote the Hessian of $s$ along $\{\d s = 0\}$.  Since we assume that $\{\d s = 0\}$ is clean, $s^{(2)}$ is nondegenerate in the fiber directions of $X_{\approx\{\d s = 0\}} \to \{\d s = 0\}$.  We can restrict $s^{(2)}$ to these fibers, take its determinant, take a square root, and take a ratio with $\dVol$ to produce a volume form $\dVol / \sqrt{|\det s^{(2)}|}$ on $\{\d s = 0\}$.  In fact, this volume form is independent of the choice of trivialization.  Using $\dVol/\sqrt{|\det s^{(2)}|}$ as a reference allows us to turn the $\CC\[\hbar\]$-valued measure $\int_{\mathrm{fiber}}f\exp\bigl(\i s/\hbar\bigr)\,\dVol$ into a function on $\{\d s = 0\}$.  By the stationary phase approximation, it begins
$$ \frac{\int_{\mathrm{fiber}}f\exp\bigl(\i s/\hbar\bigr)\,\dVol}{\dVol/\sqrt{|\det s^{(2)}|}} = f|_{\{\d s = 0\}} + O(\hbar). $$

The method of Feynman diagrams computes the higher-order coefficients of the above $\CC\[\hbar\]$-valued function.  Choose a way to identify the fiber bundle $X_{\approx\{\d s = 0\}} \to \{\d s = 0\}$ with a neighborhood of the zero-section of a vector bundle over $\{\d s = 0\}$.  Note that each fiber has a unique volume form $\dVol_{\mathrm{fiber}}$ that multiplies with the pull-back of $\dVol/\sqrt{|\det s^{(2)}|}$ to give $\dVol$; we can always choose the identification of $X_{\approx\{\d s = 0\}} \to \{\d s = 0\}$ with a vector bundle in such a way that this volume form $\dVol_{\mathrm{fiber}}$ is the pull-back of the Lebesgue measure.  Having done all this, we can define the fiberwise Taylor coefficients of $f$ and $s$.  The higher-order corrections in ``$f + O(\hbar)$'' are rational functions of these Taylor coefficients.  %In this sense, the method of Feynman diagrams is totally algebraic.

To derive these rational functions, it suffices now to restrict attention to the case when $X\cong \RR^{n}$, $\dVol$ is the Lebesgue measure, and $s$ has a nondegenerate critical point at $0\in X$, with $s(0) = 0$.  Since it suffices to consider only the Taylor expansions of $f$ and $s$ near this critical point, we take $\Oo(X)$ to be the formal power series ring $\CC_{\hbar}\[x_{1},\dots,x_{n}\]$, where $\CC_{\hbar}=\CC[\hbar^{-1}]\[\hbar\]$ denotes the field of Laurent series, and all power series rings are completed for the power-series topology.  Stokes' formula will never produce boundary terms, because we can always extend any Taylor series to a compactly-supported smooth function, and so we can control the integral with a Batalin--Vilkovisky complex.  Let $\xi_{i}$ be an odd variable corresponding to the divergence-free (for the Lebesgue measure) vector field $\frac{\partial}{\partial x_{i}}$.  As a graded algebra, our BV complex will be:
$$ \MV_{\bullet}(X) = \CC_{\hbar}\[x_{1},\dots,x_{n},\xi_{1},\dots,\xi_{n}\] $$
The $\xi_{i}$ variables are in homological degree $+1$.  

The differential, as in Example~\ref{eg.smooth}, has two terms, one corresponding to ``divergence with respect to the Lebesgue measure,'' and the other corresponding to ``contraction with $\d\bigl( \i s/\hbar \bigr)$.''  In terms of the graded power-series algebra, these differentials together are:
$$ \partial_{\BV} = \sum_{i=1}^{n}\left(\frac{\partial^{2}}{\partial x_{i}\partial \xi_{i}} + \frac\i\hbar \frac{\partial s}{\partial x_{i}}\frac{\partial}{\partial \xi_{i}}\right) $$
By construction, the map $\CC_\hbar\[x_{1},\dots,x_{n}\] \to \CC_\hbar$ taking $f$ to the formal integral $I_{\i s/\hbar}(f) = \int_{X} f\exp\bigl(\i s/\hbar\bigr)\dVol$ factors through the degree-$0$ homology of this complex, and so we are primarily interested in the following question: \emph{Given $f \in \MV_{0}(X) = \CC_{\hbar}\[x_{1},\dots,x_{n}\]$, what class does it represent in $\H_{0}\bigl(\MV_{\bullet}(X),\partial_{\BV}\bigr)$?}
The answer to this question is invariant under rescalings of $\partial_{\BV}$.  Since $\hbar$ is an infinitesimally small parameter, we will work with $-\hbar \partial_{\BV}$, as this avoids having to divide by $\hbar$.

By assumption, $s(0) = 0$ and $s$ has a critical point $0$.  We replace $s$ by its Taylor series, and make a mild change to the names of its Taylor coefficients:
$$  s(x) = \i \sum_{i,j=1}^{n} \frac12 a_{i,j} \,x_{i}x_{j} - \i \sum_{\ell\geq 3} \frac1{\ell!} \sum_{\vec\imath \in \{1,\dots,n\}^{\ell}}b^{(\ell)}_{\vec\imath}\, x_{i_{1}}x_{i_{2}}\cdots x_{i_{\ell}}$$
where the symmetric matrix $\{a_{i,j}\}_{i,j=1}^{n}$ is invertible since we supposed that $0\in X$ was nondegenerate as a critical point.  Put together, we are interested in the following differential:
$$ -\hbar \partial_{\BV} = \sum_{i,j=1}^{n} a_{i,j}\,x_{i}\,\frac{\partial}{\partial \xi_{j}} 
 - \sum_{\ell \geq 2} \frac1{\ell!} \sum_{\substack{\vec\imath \in \{1,\dots,n\}^{\ell} \\ j\in \{1,\dots,n\}}} b^{(\ell+1)}_{\vec\imath, j} \, x_{i_{1}}x_{i_{2}}\cdots x_{i_{\ell}} \, \frac{\partial}{\partial \xi_{j}}
 - \hbar \sum_{i=1}^{n} \frac{\partial^{2}}{\partial x_{i}\partial \xi_{i}} $$

The leading term of this differential removes a factor of $\xi$ in exchange for a factor of $x$.  The terms with a ``$b$'' in them are ``smaller'' in the power-series topology, in that they produce more $x$s, and the final term is ``small'' in that it produces an $\hbar$.  This suggests that we understand $-\hbar\partial_{\BV}$ as a perturbation of its leading term $\sum_{i,j=1}^{n} a_{i,j}\,x_{i}\,\frac{\partial}{\partial \xi_{j}}$.  There is a general theory of how to understand differentials as perturbations of their leading terms, which we will review in Lemma~\ref{hpl}, but this differential is sufficiently simple as to be amenable to direct analysis.

If the differential were just $\sum_{i,j=1}^{n} a_{i,j}\,x_{i}\,\frac{\partial}{\partial \xi_{j}}$, then the exact elements in degree $0$ would consist of all those power series that are first-order and higher in the $x_{i}$s.  Indeed, suppose that $f_{\vec\imath}$ is an arbitrary (not necessarily symmetric) $m$-tensor, so that the index $\vec\imath$ runs over $m$-tuples $\vec\imath \in \{1,\dots,n\}^{m}$.  Then $f_{\vec\imath}$ determines a homogeneous polynomial $\sum_{\vec\imath}f_{\vec\imath}\, x_{i_{1}}\cdots x_{i_{m}}$, which satisfies:
$$ \sum_{\vec\imath\in \{1,\dots,n\}^{m}}f_{\vec\imath}\, x_{i_{1}}\cdots x_{i_{m}} = \sum_{i,j=1}^{n} a_{i,j}\,x_{i}\,\frac{\partial}{\partial \xi_{j}} \left(\sum_{i',j' = 1}^{n} \xi_{j'} \, (a^{-1})_{i',j'} \sum_{\vec\imath\in \{1,\dots,n\}^{m-1}}f_{\vec\imath,i'}\, x_{i_{1}}\cdots x_{i_{m-1}} \right)$$
Here $(a^{-1})_{i',j'}$ denotes the $(i',j')$th entry of the inverse of the matrix with entries $a_{i,j}$.

If we instead apply $-\hbar\partial_{\BV}$ to $\sum_{i',j' = 1}^{n} \xi_{j'} \, (a^{-1})_{i',j'} \sum_{\vec\imath\in \{1,\dots,n\}^{m-1}}f_{\vec\imath,i'}\, x_{i_{1}}\cdots x_{i_{m-1}} $, we conclude that:
\begin{multline*}
\sum_{\vec\imath\in \{1,\dots,n\}^{m}}f_{\vec\imath}\, x_{i_{1}}\cdots x_{i_{m}} = 
\sum_{\substack{  i\in \{1,\dots,n\} \\ \vec\imath \in \{1,\dots,n\}^{m-1}}} f_{\vec\imath,i}\, x_{i_{1}}\cdots x_{i_{m-1}}\, 
\sum_{\ell \geq 2} \frac1{\ell!} \sum_{\substack{\vec\jmath \in \{1,\dots,n\}^{\ell} \\ j\in \{1,\dots,n\} }} (a^{-1})_{i,j}\,  b^{(\ell+1)}_{\vec\jmath, j} \, x_{j_{1}}x_{j_{2}}\cdots x_{j_{\ell}}
\\
+ \hbar \sum_{\substack{  i\in \{1,\dots,n\} \\ \vec\imath \in \{1,\dots,n\}^{m-1}}} \sum_{k=1}^{m-1} (a^{-1})_{i,i_{k}}\,f_{\vec\imath,i}\, x_{i_{1}}\cdots \widehat{x_{i_{k}}}\cdots x_{i_{m-1}}
\quad\text{modulo $\partial_{\BV}$-exact terms}
\end{multline*}
The hat denotes a term left out of the product.

This formula has a graphical interpretation.  The tensor $f_{\vec\imath}$ represents a multilinear functional of $m$ variables, each ranging over $\CC^{n}$; we can draw such an operator as a box with $m$ ordered inputs:
$$ f_{\vec\imath} = 
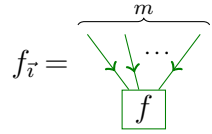
  \begin{tikzpicture}[baseline=(center),Feynman]
    \path node[draw,rectangle,inner sep=2pt] (star) {$\:f\:$} +(0,.5) coordinate (center) +(.2,.75) node {\scriptsize \dots};
    \path (star) +(-.75,1) coordinate (dot1) {}
      +(-.25,1) coordinate (dot2) 
    %  +(.5,1) coordinate (dot3) 
      +(.75,1) coordinate (dot4) ;
    \draw[inversearrow] (star) -- (dot1); \draw[inversearrow] (star) -- (dot2); %\draw[inversearrow] (star) -- (dot3);
     \draw[inversearrow] (star) -- (dot4);
    \path (dot1) ++(-.125,.125) coordinate (b1) (dot4) ++(.125,.125) coordinate (b2);
    \draw[decorate,decoration=brace,black] (b1) -- node[auto] {$\scriptstyle m$} (b2);
  \end{tikzpicture}
$$
We henceforth suppress the arrows denoting inputs and outputs to multilinear functions, using instead the convention that ``time goes down the page,'' so that inputs are at the top and outputs are at the bottom.  Tensor products are denoted by placing diagrams side-by-side, and composition (i.e.\ contraction of tensors) is denoted by connecting input and output strands.
We use an open circle~\tikz[baseline=(bottom),Feynman]{ \node[Odot] (star) {}  ++(0,-1ex) coordinate (bottom); \draw (star) -- ++(0,-1.2ex);} with one output to denote the variable vector $(x_{1},\dots,x_{n})\in \CC^{n}$, so that the polynomial $\sum_{\vec \imath} f_{\vec\imath}\, x_{i_{1}}\cdots x_{i_{m}}$ is given in pictures by:
$$ \sum_{\vec\imath\in \{1,\dots,n\}^{m}}f_{\vec\imath}\, x_{i_{1}}\cdots x_{i_{m}} = 
  \begin{tikzpicture}[baseline=(center),Feynman]
    \path node[draw,rectangle,inner sep=2pt] (star) {$\:f\:$} +(0,.5) coordinate (center) +(0,.75) node {\scriptsize \dots};
    \path (star) +(-.75,1) node[Odot] (dot1) {}
      +(-.5,1) node[Odot] (dot2) {}
      +(.5,1) node[Odot] (dot3) {}
      +(.75,1) node[Odot] (dot4) {};
    \draw (star) -- (dot1); \draw (star) -- (dot2); \draw (star) -- (dot3); \draw (star) -- (dot4);
    \path (dot1) ++(-.125,.125) coordinate (b1) (dot4) ++(.125,.125) coordinate (b2);
    \draw[decorate,decoration=brace,black] (b1) -- node[auto] {$\scriptstyle m$} (b2);
  \end{tikzpicture}
$$
We introduce a solid circle   \begin{tikzpicture}[baseline=(X),Feynman]
    \path node[dot] (O) {} ++(0pt,2pt) coordinate (X);
    \draw (O) -- ++(-11pt,11pt) +(-3pt,1pt) coordinate (l);
    \draw (O) -- ++(-6pt,11pt) ;
    \draw (O) -- ++(11pt,11pt) +(3pt,1pt) coordinate (r);
    \path (O) ++(2pt,9pt) node[color=black] {$\scriptstyle \ldots$};
    \draw[decorate,decoration=brace,black,black,color=black] (l) -- node[auto] {$\scriptstyle \ell$} (r);
  \end{tikzpicture} 
 with $\ell$ incoming edges (which are unordered) to denote  the tensor $b^{(\ell)}_{\vec\jmath}$.  Finally, we let a cap \begin{tikzpicture}[baseline=(X),Feynman]
    \path coordinate (A) ++(0pt,2pt) coordinate (X);
    \path (A);
    \path (A) ++(20pt,0) coordinate (B);
    \draw (A) .. controls +(0,15pt) and +(0,15pt) .. (B);
  \end{tikzpicture} denote the matrix $a^{-1}$ thought of as an element of $\CC^{n}\otimes \CC^{n}$.
    With this notation, the result of the computation above reads:
$$
  \begin{tikzpicture}[baseline=(center),Feynman]
    \path node[draw,rectangle,inner sep=2pt] (star) {$\:f\:$} +(0,.5) coordinate (center) +(0,.75) node {\scriptsize \dots};
    \path (star) +(-.75,1) node[Odot] (dot1) {}
      +(-.5,1) node[Odot] (dot2) {}
      +(.5,1) node[Odot] (dot3) {}
      +(.75,1) node[Odot] (dot4) {};
    \draw (star) -- (dot1); \draw (star) -- (dot2); \draw (star) -- (dot3); \draw (star) -- (dot4);
    \path (dot1) ++(-.125,.125) coordinate (b1) (dot4) ++(.125,.125) coordinate (b2);
    \draw[decorate,decoration=brace,black] (b1) -- node[auto] {$\scriptstyle m$} (b2);
  \end{tikzpicture}
  =
  \sum_{\ell=2}^\infty \frac1{\ell!}
  \begin{tikzpicture}[baseline=(center),Feynman]
    \path node[draw,rectangle,inner sep=2pt] (star) {$\:f\:$} +(0,.5) coordinate (center);
    \path (star) +(-.75,1) node[Odot] (dot1) {}
        +(-.5,1) node[Odot] (dot2) {}
        +(.5,1) node[Odot] (dot3) {}
        +(0,.75) node {\scriptsize \dots}
      ++(1.3,0) node[dot] (dot) {}
        +(-.375,1) node[Odot] (dot4) {}
        +(.375,1) node[Odot] (dot5) {}
        +(0,.75) node {\scriptsize \dots};
    \draw (star) -- (dot1); \draw (star) -- (dot2); \draw (star) -- (dot3); 
    \draw (star) ..controls +(.5,.6) and +(-.5,.6).. (dot);
    \draw (dot) -- (dot4); \draw (dot) -- (dot5);
    \path (dot1) ++(-.125,.125) coordinate (b1) (dot3) ++(.125,.125) coordinate (b2);
    \draw[decorate,decoration=brace,black] (b1) -- node[auto] {$\scriptstyle m-1$} (b2);
    \path (dot4) ++(-.125,.125) coordinate (b3) (dot5) ++(.125,.125) coordinate (b4);
    \draw[decorate,decoration=brace,black] (b3) -- node[auto] {$\scriptstyle \ell$} (b4);
  \end{tikzpicture}
  + \hbar
  \sum_{k=1}^n
  \begin{tikzpicture}[baseline=(center),Feynman]
    \path node[draw,rectangle,inner sep=2pt] (star) {$\:f\:$} +(0,.5) coordinate (center) +(0,1) node {\scriptsize \dots};
    \path (star) +(-.75,1) node[Odot] (dot1) {}
      +(-.5,1) node[Odot] (dot2) {}
      +(.5,1) node[Odot] (dot3) {};
    \draw (star) -- (dot1); \draw (star) -- (dot2); \draw (star) -- (dot3);
    \path (dot1) ++(-.125,.125) coordinate (b1) (dot3) ++(.125,.125) coordinate (b2);
    \draw[decorate,decoration=brace,black] (b1) -- node[auto] {$\scriptstyle m-2$} (b2);
    \draw (star) .. controls +(.75,.75) and +(-.25,1) .. node[anchor=south east,inner sep=0, near end] {$\scriptstyle k$}  (star);
  \end{tikzpicture}
  \quad \text{modulo $\partial_{\BV}$-exact terms.}
$$

It follows that any homogeneous polynomial $\sum_{\vec \imath} f_{\vec\imath}\, x_{i_{1}}\cdots x_{i_{m}} \in \CC_{\hbar}\[x_{1},\dots,x_{n}\]$ is cohomologous to an element of $\CC_{\hbar}$, and in fact in $\CC\[\hbar\]$.  Indeed, think of diagrams with~\tikz[baseline=(bottom),Feynman]{ \node[Odot] (star) {}  ++(0,-1ex) coordinate (bottom); \draw (star) -- ++(0,-1.2ex);}\,s as \define{many-headed Hydra}, and invite Hercules to kill one.  He chops off its right-most head, and either fuses it with one of the other heads, increasing degree in $\hbar$, or the Hydra grows at least two more new ones.  Hercules thereby produces a sequence of linear combinations of Hydra, which converges in the power-series topology.  The limit consists of all diagrams with no heads at all --- these are the only Hydra that Hercules cannot attack --- and each one appears with a factor of $\hbar^{\beta}$, where $\beta$ is the first Betti number of the diagram, and also a factor coming from the $\frac1{\ell!}$s counting the number of symmetries of the diagram.  This is the usual sum of Feynman diagrams.

We have not proved that $1\in \CC_{\hbar}$ is not $\partial_{\BV}$-exact.  One can prove this by arguing that any $\partial_{\BV}$-primitive of $1$ must satisfy a differential equation, and the solutions to this differential equation do not have Laurent-series expansions in $\hbar$.  We will leave the details to the reader.

At the end of the computation, there is the difficulty of evaluating one particular integral.  The homological calculation translates into a calculation of the ratio $I_{\i s/\hbar}(f)/I_{\i s/\hbar}(1)$, where $I_{\i s/\hbar}(f) \in \CC\[\hbar\]$ is the formal asymptotics of 
$\int_{X} f\,\exp\bigl( \i s/\hbar\bigr)\,\dVol$, and all integrals are regulated by multiplying the integrand by a compactly-supported bump function.  We have not, however, computed the asymptotics of $I_{\i s/\hbar}(1)$.  Similar algebraic techniques allow one to reduce this computation to the computation of the volume of a Gaussian distribution, which is some power of $\pi$ and not an algebraic number.

\subsection{Example (complex polynomials)} \label{eg.main}

We come now to our main example, which will occupy the remainder of this paper.  Suppose that $X = \CC^{n}$ with algebra of functions $\Oo(X) = \CC[x_{1},\dots,x_{n}]$.  Given a function $s\in \Oo(X)$, we can construct a Batalin--Vilkovisky complex by mimicking the construction from Example~\ref{eg.Feynman}.  Namely, we include (anticommuting) variables $\xi_{1},\dots,\xi_{n}$ in homological degree $1$, and construct:
$$ \MV_{\bullet}(X) = \CC[x_{1},\dots,x_{n},\xi_{1},\dots,\xi_{n}], \quad \partial_{\BV} = \sum_{i=1}^{n}\left(\frac{\partial^{2}}{\partial x_{i}\partial \xi_{i}} + \frac{\partial s}{\partial x_{i}}\frac{\partial}{\partial \xi_{i}}\right) $$
Since we are working over complex numbers and hope to work nonperturbatively, we have absorbed the factor of $\frac\i\hbar$ into the function $s$.  As in Example~\ref{eg.smooth} and Remark~\ref{remark.twisted}, the BV complex is isomorphic to some homological-degree shift of the perhaps-more-familiar twisted de Rham complex.  Our goal in Section~\ref{section.polynomials} will be to give explicit formulas for the homology classes represented by elements of $\MV_{0}(X) = \CC[x_{1},\dots,x_{n}]$, thereby giving a nonperturbative version of Feynman diagrams.

We have motivated Batalin--Vilkovisky complexes as a way to study ``oscillating'' integrals.  The choice of polynomials $f,s\in \Oo(X) = \CC[x_{1},\dots,x_{n}]$ is not enough to define an integral ``$\int_{\CC^{n}}f\,e^{s}\,\dVol$.''  We must, of course, choose a ``volume form'' $\dVol$, which we take to be the canonical holomorphic $n$-form $\dVol = \d x_{1}\cdots \d x_{n}$ on $\CC^{n}$, which suffices to determine a good notion of ``divergence'' of a vector field.  But we must also choose a {contour} for integration.  

By definition, a \define{contour} $\gamma$ is a properly immersed $n$-real-dimensional submanifold $\gamma \looparrowright \CC^{n}$.  We do not demand that $\gamma$ be compact (indeed, if $\gamma$ is compact, then $\int_{\gamma}f\,e^{s}\,\dVol = 0$), and so we must assure that the integral $\int_{\gamma}f\,e^{s}\,\dVol$ converges.  Moreover, since the BV complex encodes the integration by parts formula, we must assure that all boundary terms vanish.  Roughly speaking, a contour $\gamma$ is \define{allowable for $s$} if $s$ has very-negative real part $\Re(s)$ near the ends of $\gamma$, as then $|f\,e^{s}|$ enjoys exponential decay for any polynomial $s$.  (Recall that the \define{ends} of a non-compact space are ``the parts of the space that are outside any compact subspace,'' so that \define{allowability} is the condition  that for every $r\in \RR$, there is a compact subset $C \subseteq \gamma$ such that $\Re(s) < r$ on $\gamma\sminus C$.)  By a theorem usually named after either Stokes or Cauchy, provided convergence is maintained, homotopies of allowable contours do not change the values of integrals.  Thus we can consider the space of (linear combinations of) allowable contours up to homotopy, and in short-hand we will represent this space as a relative homology group:
$$ \{\text{allowable contours}\}/(\text{homotopy}) = \H_{n}\bigl(\CC^{n};\{\Re(s)\ll 0\} \bigr) $$

In fact, asking that $\Re(s)\ll 0$ near the ends of $\gamma$ is not sufficient to assure convergence of the integral, because $\gamma$ might have wild behavior at its ends.  But every contour with ends in $\{\Re(s)\ll 0\}$ is homotopy equivalent to one for which all integrals of the form $f\,e^{s}\,\dVol$ with polynomial $f$ converge, and  the homotopy types of these spaces of contours are equivalent~\cite{MR579748,MR713258}.

$$ 
\begin{tikzpicture}
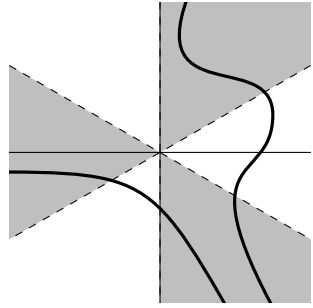
  \node at (0,-2.5) {A basis for $\{$allowable contours$\}/($homotopy$)$ when $s = x^{3}$.};
  \clip (-2,-2) rectangle (2,2);
  \draw[dashed,fill=gray!50] (30:4) -- (0,0) -- (90:4);
  \draw[dashed,fill=gray!50] (150:4) -- (0,0) -- (210:4);
  \draw[dashed,fill=gray!50] (270:4) -- (0,0) -- (330:4);
  \draw (-3,0) -- (3,0); \draw (0,-3) -- (0,3);
  \draw[very thick] (185:3) .. controls +(0:3) and +(120:3) .. (295:3);
  \draw[very thick] (305:3) .. controls +(120:3) and +(270:1) .. (1.5,.5) .. controls +(90:1) and +(240:3) .. (75:3);
\end{tikzpicture} $$

The usual way to study the relative homology group $\H_{n}\bigl(\CC^{n};\{\Re(s)\ll 0\} \bigr)$ is through the theory of Lefschetz thimbles.  This theory is well-developed, and so we will give only a summary account, referring the interested reader to~\cite{MR713258,MR2809462}.  Suppose that $s$ has only isolated critical points with no critical point at $\infty$, and suppose furthermore that the Hessian of $s$ is nondegenerate at each critical point.  (These conditions hold for generic polynomials $s$.)  Then $\Re(s)$ is a Morse function on $\CC^{n} = \RR^{2n}$ for which all critical points have Morse index $n$.  The \define{Lefschetz thimble} for a critical point $p$ of $s$ is the $n$-dimensional disk of points in $\CC^{n}$ that can be reached by downward gradient flow from $p$.  By general Morse theory, the Lefschetz thimbles form a $\ZZ$-basis of the relative homology group $\H_{n}\bigl(\CC^{n};\{\Re(s)\ll 0\} \bigr)$, and moreover witness that:
$$ \text{For generic $s$, } \H_{k}\bigl(\CC^{n};\{\Re(s)\ll 0\} \bigr) = \begin{cases} 0, & k\neq n,
  \\
\ZZ^{\#\{\d s = 0\}} , & k = n,
\end{cases}
$$
where $\#\{\d s = 0\}$ is the number of critical points of $s$.  If $s$ is generic of maximum total degree $d$ in $n$ variables, then $\#\{\d s = 0\} = (d-1)^{n}$.

Another important situation is when $s$ is required to be homogenous.  Suppose that $s \in \Oo(X) = \CC[x_{1},\dots,x_{n}]$ is homogeneous of total degree $d$, and that the corresponding projective hypersurface $\{s = 0\} \subseteq \CC\PP^{n-1}$ is smooth.  Then  the critical locus $\{\d s = 0\}$ consists of the origin with multiplicity $(d-1)^{n}$.  The relative homology group $\H_{n}\bigl(\CC^{n};\{\Re(s)\ll 0\} \bigr)$ in this case can be seen directly to be free on $(d-1)^{n}$ generators (see e.g.~\cite{MO90404}).  

These results suggest that there is in general a close relationship between the relative homology group $\H_{n}\bigl(\CC^{n};\{\Re(s)\ll 0\} \bigr)$ and the critical locus $\{\d s = 0\}$.  There are also many close relationships between the BV complex and the algebra of functions $\Oo(\{\d s = 0\})$ on the critical locus.  Indeed, as we will construct in Theorem~\ref{thm.main}, in many cases there are isomorphisms $\H_{0}\bigl(\BV_{\bullet}(X,s),\partial_{\BV}\bigr) \cong \Oo(\{\d s = 0\})$, and of course $\dim \Oo(\{\d s = 0\}) = \#\{\d s = 0\}$, counted with multiplicity.
The BV complex is also related to other aspects of the geometry of the polynomial $s: \CC^{n}\to \CC$; for example, after a grading shift, it is quasiisomorphic to the reduced algebraic de Rham cohomology of the fiber of $s$ over the scheme-theoretic generic point in~$\CC$~\cite{MR1214449}, and in particular the BV homology is finite-dimensional for arbitrary~$s$.

All together, we have a topologically-defined space $\H_{n}\bigl(\CC^{n};\{\Re(s)\ll 0\} \bigr)$ of contours $\gamma$, and an algebraically-defined space $\H_{0}\bigl( \MV_{\bullet}(X),\partial_{\BV}\bigr)$ of observables $f$, which have a canonical analytically-defined pairing $(\gamma,f) \mapsto \int_{\gamma}f\,e^{s}\,\dVol \in \CC$.  
This pairing is well-studied and has many generalizations.  In particular, it is known to be a perfect pairing of finite-dimensional vector spaces~\cite{MR2443034}.

\section{A non-asymptotic analog of Feynman diagrams} \label{section.polynomials}

In this section we study homological-algebraic aspects of the Batalin--Vilkovisky complex introduced in Example~\ref{eg.main}.  We begin by reviewing notation and definitions from Section~\ref{section.BVmotivation}.  We then state our main Theorem~\ref{thm.main} constructing explicit isomorphisms between the quantum and classical BV homologies; the formulas presenting such isomorphisms are a non-asymptotic version of the method of Feynman diagrams, and allow the ``higher modes'' of any ``oscillating'' integral to be integrated out in a totally algebraic fashion, resulting in an integral over the scheme-theoretic critical locus.  Theorem~\ref{thm.basis} and Corollary~\ref{my.thm4} give a complete list of integrals (for generic action $s$) that cannot be computed algebraically; all other integrals are determined by these  and pure algebra.

\subsection{Definitions and notation}

%As analytic aspect of integration will not appear in the remainder of this paper, we have many choices of ring over which to work.  Nevertheless, we will continue to call our ground ring $\CC$, and occasionally we will use that $\CC$ is a field of characteristic $0$.  
We fix a positive integer $n$, and write $\Oo(X) = \CC[x_{1},\dots,x_{n}]$ for the polynomial ring in $n$ variables.  We set $\MV_{\bullet}(X) = \CC[x_{1},\dots,x_{n},\xi_{1},\dots,\xi_{n}]$ to be the $\ZZ$-graded ring of antisymmetric polynomial multivector fields on $X = \AA^{n}$, where the $\xi$ variables have homological degree $+1$ and anticommute with each other.  

Choose $s\in \Oo(X)$.  We will denote by $d$ the maximal homogeneous degree of $s$.  Write $s = \sum_{k=0}^{d} s^{(k)}$ where each $s^{(k)}$ is homogeneous of degree $k$; then $s^{(d)}$ is the \define{top part} of $s$.  Every homogeneous polynomial in $n$ variables defines a hypersurface in $\CC\PP^{n-1}$.  We say that $s$ has \define{nonsingular top part} if the hypersurface defined by $s^{(d)}$ is smooth, or equivalently if the discriminant of $s^{(d)}$ is non-zero.

The \define{scheme-theoretic critical locus} of $s$ is the subscheme of $X$ with ring of functions $\Oo(\{\d s = 0\}) = \Oo(X) / \bigl( \sum_{i}\frac{\partial s}{\partial x_{i}}\Oo(X)\bigr)$, which is also known as the \define{Jacobian ring} of $s$.  This ring of functions appears as the degree-$0$ homology of the \define{Koszul resolution} of $\{\d s = 0\}$, which is the differential graded algebra with underlying graded algebra $\MV_{\bullet}(X)$ and differential
$$ \partial_{\cl} = \sum_{i=1}^{n} \frac{\partial s}{\partial x_{i}} \frac{\partial}{\partial \xi_{i}}. $$
The complex $\bigl(\MV_{\bullet}(X),\partial_{\cl}\bigr)$ is also known as the \define{classical BV complex for $s$} and as the \define{derived critical locus of $s$}.

The \define{divergence} operator corresponding to the volume form $\dVol = \d x_{1}\cdots \d x_{n}$ is the differential on $\MV_{\bullet}(X)$ given by:
$$ \div = \sum_{i=1}^{n} \frac{\partial^{2}}{\partial x_{i}\partial \xi_{i}}$$
The \define{(quantum) BV complex for $s$} is:
$$ \BV_{\bullet}(X,s) = \bigl( \MV_{\bullet}(X), \partial_{\BV} = \partial_{\cl} + \div\bigr). $$

\subsection{Remark (complete intersections)} \label{remark.serre}

The critical locus $\{\d s = 0\}$ is \define{zero-dimensional} if $\dim \Oo(\{\d s = 0\}) < \infty$, in which case we define $\#\{\d s = 0\} = \dim \Oo(\{\d s =0 \})$.
If the critical locus is zero-dimensional and there are no critical points at infinity, then the critical locus is a complete intersection and a theorem of Serre's implies that  the Koszul resolution $\bigl(\MV_{\bullet}(X),\partial_{\cl}\bigr)$ has homology entirely in degree zero.

If $s$ is of degree $d$ and has nonsingular top part, then the critical locus $\{\d s = 0\}$ is zero-dimensional and Bezout's theorem implies that $\#\{\d s = 0\} = (d-1)^{n}$.  Indeed, the top part $s^{(d)}$ cannot have critical points at infinity if it is nonsingular, and so $\{\d s^{(d)} = 0\}$ is a complete intersection with $\#\{\d s^{(d)} = 0\} = (d-1)^{n}$; the corresponding statement for $s$ will follow, for example, from our Proof~\ref{proof.main} of Theorem~\ref{thm.main}.

\subsection{Remark (strategy to study the quantum BV complex)}

The classical and quantum BV complexes for $s$ are closely related.  For example, by filtering $\Oo(X)$ by polynomial degree, one can prove from a spectral sequence argument that there exists a differential $\tilde\partial$ on the classical BV homology $\H_{\bullet}\bigl(\MV_{\bullet}(X),\partial_{\cl}\bigr)$ and isomorphisms
$$ \H_{\bullet}\bigl( \H_{\bullet}\bigl(\MV_{\bullet}(X),\partial_{\cl}\bigr), \tilde\partial\bigr) \cong \H_{\bullet}\bigl(\MV_{\bullet}(X),\partial_{\BV}\bigr). $$
In particular, if the critical locus $\{\d s = 0\}$ is zero-dimensional, then the classical BV homology is concentrated in degree $0$ and so $\tilde\partial$ must vanish.  Therefore there exist isomorphisms $\H_{0}\bigl(\MV_{\bullet},\partial_{\BV}\bigr) \cong \H_{0}\bigl(\MV_{\bullet},\partial_{\cl}\bigr) = \Oo(\{\d s = 0\})$.

If $\CC$ were replaced by a non-semisimple ring, we would not immediately be able to guarantee such an isomorphism.  The correct statement filters each homology group by declaring that a class is in the $k$th filtered piece if it is represented by a degree-$k$ polynomial, and then compares $\H_{\bullet}\bigl( \gr \H_{\bullet}\bigl(\MV_{\bullet}(X),\partial_{\cl}\bigr), \tilde\partial\bigr)$ with $\gr \H_{\bullet}\bigl(\MV_{\bullet}(X),\partial_{\BV}\bigr)$, where $\gr$ denotes the associated-graded functor from filtered modules to graded modules.  Over a field, or more generally over a semisimple ring, there are always non-canonical isomorphisms between filtered modules and their associated gradeds.

The problem with the usual spectral-sequence approach is that it does not pick out explicit formulas.  The differential $\tilde\partial$ is not canonical, because the identification of a filtered vector space with its associated graded is not canonical.  Moreover, the classical and quantum BV complexes each have algebraic structure which is lost by the spectral sequence.

Most importantly, our goal is to produce explicit formulas for integrals against $e^{s}$, and this goal translates into the problem of producing an explicit basis of $\H_{0}\bigl(\MV_{\bullet}(X),\partial_{\BV}\bigr)$ and giving an explicit formula for the map taking each element of $\MV_{0}(X) = \Oo(X)$ to its class in the quantum BV homology.

Our strategy to construct such explicit formulas will be to use a formula-full analog of spectral sequences called the Homotopy Perturbation Lemma~\ref{hpl}.  We will focus on the simplest case when $s$ has nonsingular top part; it is already rich enough to provide an illustrative example.

\subsection{Theorem (integrating out the higher modes)} \label{thm.main}
We will prove the following claims in~\ref{proof.main}:

Let $s$ be a degree-$d$ polynomial with nonsingular top part.  Filter $\Oo(\{\d s = 0\})$ by declaring that a class $[f]\in \Oo(\{\d s = 0 \})$ is of degree at most $k$ if it is the restriction to $\{\d s = 0\}$ for a polynomial $f\in \Oo(X)$ of total degree at most $k$.  Denote the map that restricts functions to the critical locus by $\tau: \Oo(X) \to \Oo(\{\d s = 0\})$.  Also denote the restriction map from zero-chains to their BV-homology classes by $\tilde \tau : \Oo(X) = \MV_{0}(X) \to \H_{0}\bigl(\MV_{\bullet}(X),\partial_{\BV}\bigr)$.

Let $\varphi : \Oo(\{\d s = 0\}) \to \Oo(X)$ be any linear  splitting of $\tau$ which is filtration-non-increasing.  %Said another way, if $[f]$ is a function on $\{\d s = 0\}$ of degree at most $k$, then $\varphi([f])$ should be an extension of $[f]$ to all of $X$ which is also of degree at most $k$.
Then there exists a unique isomorphism $\Oo(\{\d s = 0\}) \cong \H_{0}\bigl(\MV_{\bullet}(X),\partial_{\BV}\bigr)$ of vector spaces such that $\varphi$ also splits $\tilde\tau$.  The homology group $\H_{0}\bigl(\MV_{\bullet}(X),\partial_{\BV}\bigr)$ can be similarly filtered by declaring that the classes of degree at most $k$ are the restrictions of functions of degree at most $k$; with this filtration, the unique isomorphism is an isomorphism of filtered vector spaces.
$$ \begin{tikzpicture}
  \path (0,0) node (Cx) {$\Oo(X)$} 
    (0,-2) node (cl) {$\Oo(\{\d s = 0\})$}
    (4,-2) node (q) {$\H_{0}\bigl(\MV_{\bullet}(X),\partial_{\BV}\bigr)$};
  \draw[->] (Cx) -- node[auto,swap] {$\scriptstyle \tau$} (cl);
  \draw[->] (Cx) -- node[auto] {$\scriptstyle \tilde\tau$} (q);
  \draw[->%, blue
  ] (cl) .. controls +(.5,1) and +(.5,-1).. node[auto,swap] {$\scriptstyle \varphi$} (Cx);
  \draw[%green!50!
  black] (cl) -- node[auto] {$\sim$} (q);
  \path
  (6,-.3) node[anchor=west] {
  %\color{blue}
  For every filtered
  \tikz[baseline=(o)] \draw[%blue,
  ->] coordinate (o) ++(0,-3pt) coordinate (b) .. controls +(3pt,6pt) and +(3pt,-6pt).. ++(0,12pt);
   such that 
  $\tikz[baseline=(o)] \draw[->] coordinate (o) ++(0,-2pt)  .. controls +(3pt,6pt) and +(3pt,-6pt).. ++(0,11pt) -- ++(-2pt,0)  -- ++(0,-11pt); = \id$% 
   ,}
  ++(0,-18pt) node[anchor=west] {
  %\color{green!50!black}
  there exists a unique
  $\simeq$ such that
  $
   \tikz[baseline=(o)] \draw[->] coordinate (o) ++(0,-2pt) -- ++(-9pt,0) .. controls +(3pt,6pt) and +(3pt,-6pt).. ++(0,11pt) -- ++(3pt,0) -- ++(10pt,-11pt);   
   = \id
  $. 
  };
\end{tikzpicture}
$$

Recall that $\Oo(\{\d s = 0\})$ appears as the degree-$0$ homology of the classical BV complex, i.e.\ $\Oo(\{\d s = 0\}) = \MV_{0}(X) / \partial_{\cl}\bigl(\MV_{1}(X)\bigr)$.  We give $\MV_{1}(X) = \bigoplus_{i}\Oo(X)\xi_{i}$ a filtration generated by the polynomial degree in $\Oo(X)$ and the declaration that each $\xi_{i}$ is of degree $d-1$.  
Choose any filtration-non-increasing \define{homotopy} $\eta: \MV_{0}(X) \to \MV_{1}(X)$ such that $\partial_{\cl}\circ \eta = \id - \varphi\circ \tau$.  
(We will prove that such $\eta$ exists.)
Then the composition ${\div} \circ \eta = \sum_{i} \frac{\partial^{2}}{\partial \xi_{i}\partial x_{i}}\circ \eta : \MV_{0}(X) \to \MV_{0}(X)$ acts locally nilpotently, as it lowers filtration-degree by at least $d$, and so any power series in ${\div} \circ \eta$ converges.  With respect to the isomorphism $\Oo(\{\d s = 0\}) \cong \H_{0}\bigl(\MV_{\bullet}(X),\partial_{\BV}\bigr)$ determined by $\varphi$, the restriction map $\tilde\tau : \MV_{0}(X) \to \H_{0}\bigl(\MV_{\bullet}(X),\partial_{\BV}\bigr)$ is given by the formula:
$$ \tilde\tau(f) = \tau  (\id - {\div}\circ \eta)^{-1}(f) = \sum_{k=0}^{\deg(f)/d} \tau  \bigl({\div} \circ \eta\bigr)^{k}(f). $$

\subsection{Remark (inaccurate geometric interpretation)}

The critical locus $\{\d s = 0\}$ is a subscheme of $X = \AA^{n}$.  In Example~\ref{eg.Feynman}, after restricting attention $X \leadsto X_{\approx\{\d s = 0\}}$ to a tubular neighborhood of the critical locus, we chose a fibration of $X_{\approx\{\d s = 0\}}$ over $\{\d s = 0\}$, and integrated over the fibers to produce a function on $\{\d s = 0\}$.  In our present setting of polynomials  and schemes, we cannot retract $X$ to the critical locus in any geometric way, so there are no fibers to integrate  over.  

Nevertheless, that is how Theorem~\ref{thm.main} should be interpreted.  The choice of linear map $\varphi$ acts as if it were a retraction of $X$ onto $\{\d s = 0\}$.
 The composition $\tilde\tau : \Oo(X) \to \H_{0}\bigl(\MV_{\bullet}(X),\partial_{\BV}\bigr) \cong \Oo(\{\d s = 0\})$ is the map that takes a function on $X$ and ``integrates it over the fibers'' of this $\varphi$ to produce a function on the critical locus $\{\d s = 0\}$.
Comparing further with the Feynman diagrams in Example~\ref{eg.Feynman}, the choice of homotopy $\eta$ plays the role of a propagator, and the operation ${\div}\circ \eta$ corresponds to playing one round of ``many-headed Hydra.''

\subsection{Example (Wick's formula)}\label{eg.wick}

An important example is when $s$ is quadratic with nonsingular top part, although this example is too simple to illustrate many of the phenomena present when $\deg(s) > 2$.  The critical locus $\{\d s = 0\}$ is then a single point, and there is a unique filtration-non-increasing map $\varphi: \CC =  \Oo(\{\d s = 0 \}) \to \Oo(X) = \CC[x_{1},\dots,x_{n}]$, and it happens to be an algebra homomorphism.

Expand $s$ in coordinates:
$$ s = \sum_{i,j=1}^{n} \frac12\, s^{(2)}_{i,j}\, x_{i}x_{j} + \sum_{i=1}^{n} s^{(1)}_{i}\,x_{i} + s^{(0)} $$
Then $\{\d s = 0\}$ consists of the point at coordinates $x_{i} = - \sum_{j} \bigl(s^{(2)}\bigr)^{-1}_{i,j}\, s^{(1)}_{j}$ where $\bigl(s^{(2)}\bigr)^{-1}_{i,j}$ is the $(i,j)$th entry in the inverse matrix to $s^{(2)}_{i,j}$.  The classical BV complex has differential:
$$ \partial_{\cl} = \sum_{i,j=1}^{n} s^{(2)}_{i,j}\,x_{i}\,\frac{\partial}{\partial \xi_{j}} + \sum_{i=1}^{n} s^{(1)}_{i}\,\frac{\partial}{\partial \xi_{i}} $$

Because of the simplicity of the example, there is  a unique filtration-non-increasing homotopy $\eta : \Oo(X) \to \MV_{1}(X)$ satisfying $\partial_{\cl}\circ \eta = \id - \varphi\circ \tau$.  It is easiest to describe after changing coordinates to $y_{i} = x_{i} + \sum_{j} \bigl(s^{(2)}\bigr)^{-1}_{i,j}\, s^{(1)}_{j}$.  Then $\partial_{\cl} = \sum_{i,j} s^{(2)}_{i,j}\,y_{i}\,\frac{\partial}{\partial \xi_{j}}$, and:
$$ \eta(f) = \begin{cases}
 0 & \text{if $f$ is constant} \\
 \frac1\ell \sum_{i,j} \bigl(s^{(2)}\bigr)^{-1}_{i,j}\,\xi_{i}\,\frac{\partial f}{\partial y_{j}}
 & \text{if $f$ is homogeneous in the $y$ variables of degree $\ell>0$}.
\end{cases}
$$
Composing with $\div = \sum_{i} \frac{\partial^{2}}{\partial x_{i}\partial \xi_{i}} = \sum_{i}\frac{\partial^{2}}{\partial y_{i}\partial \xi_{i}}$ gives:
$$ {\div}\circ\eta  = \frac1\ell \sum_{i,j=1}^{n}\bigl(s^{(2)}\bigr)^{-1}_{i,j}\,\frac{\partial^{2}f}{\partial y_{i}\partial y_{j}} \text{ on functions that are homogeneous in  $y$  of degree $\ell>0$.}$$

In particular, if $f$ is homogeneous of degree $\ell$ in the $y$ variables, then $\div(\eta(f))$ is homogeneous of degree $\ell-2$.  Since $\tau$ evaluates a function at $y=0$, we see that $\tau \circ \bigl({\div}\circ \eta\bigr)^{k} (f) = 0$ for every $k$ if $f$ is homogeneous in $y$ of odd degree.  If $f$ is homogeneous in $y$ of even degree $\ell$, then $\tau \circ \bigl({\div}\circ \eta\bigr)^{k} (f) \neq 0$ only when $k = \ell/2$, in which case:
$$ \tau \circ \bigl({\div}\circ \eta\bigr)^{k} (f) =  \frac1{2k(2k-2)(2k-4)\cdots 2} \left( \sum_{i,j=1}^{n}\bigl(s^{(2)}\bigr)^{-1}_{i,j}\,\frac{\partial^{2}}{\partial y_{i}\partial y_{j}}\right)^{k}f
$$

Summing over the possible values of $k$, and restoring to the $x$ variables, this implies the following version of Wick's formula:
$$ \tilde\tau(f) = \left.\exp\left( \frac12 \sum_{i,j=1}^{n}\bigl(s^{(2)}\bigr)^{-1}_{i,j}\,\frac{\partial^{2}}{\partial x_{i}\partial x_{j}}\right) f\right|_{x_{i} = - \sum_{j} (s^{(2)})^{-1}_{i,j}\, s^{(1)}_{j}}$$
``Wick's formula'' is originally due to Isserlis \cite{Isserlis:1918kx}.

\subsection{Remark (the inverse problem)}

Fix a quadratic $s$ with nonsingular top part and a contour and let $I_{s}(f) = \int f\,e^{s}$.  Example~\ref{eg.wick} says that $I_{s}(f) = \tilde\tau(f)\,I_{s}(1)$, where $\tilde\tau(f)$ is determined algebraically in terms of $f$ and $s$.  Moreover, it implies:
\begin{gather*}
  \frac{I_{s}(x_{i})}{I(1)} = - \sum_{j} (s^{(2)})^{-1}_{i,j}\, s^{(1)}_{j} \\
  \frac{I_{s}(y_{i}y_{j})}{I_{s}(1)} = (s^{(2)})^{-1}_{i,j} \text{ where } y_{i} = x_{i} - \frac{I_{s}(x_{i})}{I_{s}(1)}
\end{gather*}
Then the coefficients of $s$ are rational functions in these values, and
 Example~\ref{eg.wick} implies that $I_{s}(f)$ is a polynomial in the values of $I_{s}(1)$, $I_{s}(x_{i})$, and $I_{s}(x_{i}x_{j})$ just for those variables $x_{i},x_{j}$ appearing in $f$.

When $s$ is not quadratic, the ability to ignore variables that do not appear in $f$ probably is not possible.  But we can ask whether there exists a finite list of functions $f$ such that the values of $I_{s}(f)$ for that list, along with \emph{a priori} knowledge of the degree of $s$, is enough to determine $s$ and the contour.  A partial result in this direction, which we will prove in Proof~\ref{proof.basis}, is the following:

\subsection{Theorem (towards a higher Wick's formula)} \label{thm.basis}

Let $s\in \Oo(X) = \CC[x_{1},\dots,x_{n}]$ have maximum total degree $\deg(s) = d$, and suppose that the top part $s^{(d)}$ is generic.  Then the quantum BV homology $\H_{0}\bigl(\MV_{\bullet}(X),\partial_{\BV}\bigr)$ has a basis consisting of the classes of the $(d-1)^{n}$ monomials $x_{1}^{m_{1}}\cdots x_{n}^{m_{n}}$ for which $m_{i} < d-1$ for all $i=1,\dots,n$. 

 With respect to this basis and the monomial basis of $\Oo(X)$, in Corollary~\ref{my.thm4} we will give an explicit formula for the restriction map $\tilde\tau$ as a rational function of the coefficients of $s$ when $s = s^{(d)}$ is homogeneous. 

There are $(d-1)^{n}$ degrees of freedom in the choice of contour, and $\binom{n+d}d$ degrees of freedom in the choice of $s$.  Since $\tilde\tau$ depends rationally on the coefficients of $s$, each function $f\in \Oo(X)$ determines an explicit rational equation satisfied by these $(d-1)^{n} + \binom{n+d}d$ variables, of the following form: for each $\vec m\in \{0,\dots,d-2\}^{n}$, the map $\tilde\tau$ produces a coefficient $\tilde\tau(f)_{\vec m}$ such that $\tilde\tau(f) = \sum_{\vec m} \tilde\tau(f)_{\vec m}\, [x_{1}^{m_{1}}\cdots x_{n}^{m_{n}}]$; to $f$ we associate the equation $I(\tilde\tau(f)) = \sum_{\vec m} \tilde\tau(f)_{\vec m}\,I(x_{1}^{m_{1}}\cdots x_{n}^{m_{n}})$.
  One  expects, therefore, that the system constructed by testing $(d-1)^{n} + \binom{n+d}d$ functions $f$ has a finite number of solutions, and these solutions are distinguished by testing one more function.

\subsection{Remark}  There are  more general results available concerning bases for Jacobian rings like $\H_{0}\bigl(\MV_{\bullet}(X),\partial_{\cl}\bigr)$ (e.g.~\cite{MR0419433,MR2205838,MR2168243,MR2229875}), and by Theorem~\ref{thm.main}  such results translate directly to the quantum BV homology.  What we would like to emphasize in this paper are the techniques used to prove Theorems~\ref{thm.main} and~\ref{thm.basis}, as they can be generalized to physically-interesting infinite-dimensional settings whereas classical singularity theoretic techniques seem more finite-dimensionally bound.

We turn now to the main ingredient in Proofs~\ref{proof.main} and~\ref{proof.basis}:

\subsection{The Homological Perturbation Lemma} \label{hpl}

Homological perturbation has a long history starting in the 1960s and described in detail in~\cite{MR2762538}.
The following definitions and result apply in any  category enriched over abelian groups.  We will intentionally use many of the same letters ($\varphi,\tau,\eta,\dots$) as we used in the statement of Theorem~\ref{thm.main}.

A \define{retraction} consists of two chain complexes $(V_{\bullet},\partial_{V})$ and $(H_{\bullet},\partial_{H})$, chain maps $\varphi: H \to V$ and $\tau: V \to H$, and a \define{homotopy} (of homological degree $+1$) $\eta : V \to V$.  These maps are required to satisfy that $\tau \circ \varphi = \id_{H}$ and $\varphi \circ \tau = {\id_{V}} + [\partial_{V},\eta]$.
  $$ \begin{tikzpicture}[baseline=(L.base),anchor=base,->,auto,swap]
      \path node (L) {$(H_{\bullet},\partial_{H})$} ++(3,0) node (M) {$(V_{\bullet},\partial_{V})$} (M.mid) +(0,.075) coordinate (raise) +(0,-.075) coordinate (lower);
      \draw (L.east |- lower) to node {$\scriptstyle \varphi$} (M.west |- lower);
      \draw (M.west |- raise) to node {$\scriptstyle \tau$} (L.east |- raise);
      \draw (M.south east) ..controls +(1,-.5) and +(1,.5) .. node {$\scriptstyle \eta$} (M.north east);
    \end{tikzpicture}
    \quad\quad
    \begin{array}{l}
     \tau \circ \varphi = {\id_{H}} \\ \varphi\circ\tau = {\id_{V}} - [\partial_{V},\eta]
    \end{array}
  $$
  It follows that $\varphi$ and $\tau$ are quasi-isomorphisms.  The commutator is to be understood with the appropriate signs: since $\partial_{V}$ is of homological degree $-1$ and $\eta$ is of degree $+1$, both of which are odd, we have $[\partial_{V},\eta] = \partial_{V}\circ \eta + \eta\circ\partial_{V}$.  It is standard but unnecessary to also impose \define{side conditions} that $\eta^{2} = 0$, $\eta \circ \varphi = 0$, and $\tau \circ \eta = 0$.

A \define{deformation} of a chain complex $(V_{\bullet},\partial_{V})$ is a degree-$(-1)$ map $\delta: V \to V$ such that $(\partial_{V} + \delta)^{2} = 0$.  Equivalently, $\delta$ is a \define{Maurer--Cartan element} of $\operatorname{End}(V)$.  A deformation $\delta$ is \define{small} with respect to a given retraction $(V_{\bullet},\partial_{V},H_{\bullet},\partial_{H},\varphi,\tau,\eta)$ if the degree-$0$ map $(\id_{V} - \delta\eta)$ is invertible.  Note that then $({\id_{V}} - \eta\delta)^{-1} = {\id_{V}} + \eta({\id_{V}} - \delta\eta)^{-1}\delta$ also exists.

Suppose we are given a retraction $(V_{\bullet},\partial_{V},H_{\bullet},\partial_{H},\varphi,\tau,\eta)$ and a small deformation $\delta$ of $(V_{\bullet},\partial_{V})$.  Then the deformed complex $(V_{\bullet},\partial_{V} + \delta)$ is part of a deformed retraction:
  $$
  \begin{tikzpicture}[baseline=(L.base),anchor=base,->,auto,swap]
      \path node (L) {$\bigl(H_{\bullet},\tilde\partial_{H} = \partial_{H} + \tau \circ ({\id} - \delta \eta)^{-1}\delta \circ \varphi\bigr)$} ++(8,0) node (M) {$(V_{\bullet},\partial_{V} +\delta)$} (M.mid) +(0,.075) coordinate (raise) +(0,-.075) coordinate (lower);
      \draw (L.east |- lower) to node {$\scriptstyle \tilde\varphi \,=\, ({\id} - \eta\delta)^{-1}\circ \varphi$} (M.west |- lower);
      \draw (M.west |- raise) to node {$\scriptstyle \tilde\tau \,=\, \tau \circ ({\id} - \delta\eta)^{-1}$} (L.east |- raise);
      \draw (M.south east) ..controls +(1,-.5) and +(1,.5) .. node {$\scriptstyle \tilde\eta \,=\,\eta ({\id} - \delta \eta)^{-1}$} (M.north east);
  \end{tikzpicture}
  $$
The graded vector spaces $V_{\bullet}$ and $H_{\bullet}$ do not change, but their differentials do.

The proof consists simply of checking some equations, and we leave it to the reader;  
a particularly good write-up of is~\cite{Crainic04}.
%In fact, at the cost of working harder at the proof one can even drop the condition that $\tau\circ\varphi = \id_{H}$, replacing it only with the condition that $\tau$ and $\varphi$ be quasi-isomorphisms~\cite{Crainic04}, but we will not need such generality.

%\subsection{Remark (Maurer--Cartan element as an algebraic structure)}  The Homological Perturbation Lemma~\ref{hpl}  is an example of a much more general homotopical perturbation theory allowing to move (strongly homotopy) algebraic structures across quasi-isomorphisms.  In the case of the Lemma~\ref{hpl}, the algebraic structure to be moved is ``the choice of a Maurer--Cartan element'': we turn $\delta$ into the Maurer--Cartan element $\tau \circ ({\id} - \delta \eta)^{-1}\delta \circ \varphi$ on $(H_{\bullet},\partial_{H})$.

\subsection{Corollary (Betti numbers are upper-semicontinuous and Euler characteristic is locally constant)}  Work in the category of vector spaces.  Given a complex $(V_{\bullet},\partial_{V})$, set $H_{\bullet} = \H_{\bullet}(V_{\bullet},\partial_{V})$ its homology, with $\partial_{H} = 0$.  Then it is possible to choose a retraction of $V_{\bullet}$ onto~$H_{\bullet}$.

It follows that the dimension of the $j$th homology group of $(V_{\bullet},\partial_{V})$ is an upper semi-continuous function of $\partial_{V}$.  Suppose that $\dim H_{\bullet} < \infty$.  Then the Euler characteristic of $(V_{\bullet},\partial_{V})$ is constant under small deformations of $\partial_{V}$.

In particular, if $\H_{\bullet}(V_{\bullet},\partial_{V})$ is supported entirely in degree $0$, then (up to isomorphism) it cannot change under small deformations of $\partial_{V}$. % This holds more generally if $(V_{\bullet},\partial_{V})$ has no consecutive non-zero homology groups.
 \label{cor.fact.hpl}

\subsection{Remark (often $\tilde \tau$ is independent of $\eta$)} Suppose that $H_{\bullet}$ is supported entirely in homological degree $0$ and that $V_{\bullet}$ is supported entirely in nonnegative degrees.  
Then the map $\tau$ is entirely determined by the map $\varphi$ and the condition that $\tau \circ \varphi = \id_{H}$.  Usually the homotopy $\eta$ is not  uniquely determined.  

In this situation, for any small deformation $\delta$, we have $\tilde\varphi = \varphi$, as $\delta\circ \varphi = 0$.  Our ultimate goal will be to compute the deformed $\tilde\tau$.  To write explicit formulas, we must choose a homotopy $\eta$, but this choice doesn't matter provided it can be made such that $(\id_{V} - \delta\eta)$ is invertible. \label{rmk.fact.hpl2}

\subsection{Proof of Theorem~\ref{thm.main}} \label{proof.main}

Let $V_{\bullet} = \MV_{\bullet}(X)$ and $H_{\bullet} = \H_{\bullet}\bigl(\MV_{\bullet}(X),\partial_{\cl}\bigr)$.  Since $s$ has nonsingular top part, the critical locus $\{\d s = 0\}$ is a complete intersection, and therefore $H_{\bullet}$ is concentrated in degree $0$ by Remark~\ref{remark.serre}.  Therefore the natural projection $\tau: V_{0}\to H_{0}$ extends by zero to a chain map $\tau: V_{\bullet} \to H_{\bullet}$.  The choice of $\varphi : H_{0}\to V_{0}$ in the statement of Theorem~\ref{thm.main} extends by zero to a chain map $\varphi: H_{\bullet}\to V_{\bullet}$.  We consider deforming $\partial_{V} = \partial_{\cl}$ by $\delta = \div$.  We can thus apply Lemma~\ref{hpl} and Remark~\ref{rmk.fact.hpl2} provided a homotopy $\eta: V_{\bullet}\to V_{\bullet+1}$ can be found such that $(\id_{V} - \delta\eta)$ is invertible, and the conclusions of Theorem~\ref{thm.main} would follow.  

We give the algebra $\MV_{\bullet}(X) = \CC[x_{1},\dots,x_{n},\xi_{1},\dots,\xi_{n}]$ a (bosonic) grading by declaring that $\deg(x_{i}) = 1$ and $\deg(\xi_{i}) = d-1$.  Consider the differential $\partial_{(d)} = \sum_{i} \frac{\partial s^{(d)}}{\partial x_{i}}\frac{\partial}{\partial \xi_{i}}$ on $\MV_{\bullet}(X)$.  Since $s^{(d)}$ is nonsingular, by Remark~\ref{remark.serre} the homology $\H_{\bullet}\bigl(\MV_{\bullet}(X),\partial_{(d)}\bigr)$ is concentrated in degree $0$, where its homology is $\Oo(\{\d s^{(d)} = 0\})$.  Moreover, $\partial_{(d)}$ has weight zero with respect to this new grading on $\MV_{\bullet}(X)$, and $\bigl(\MV_{\bullet}(X),\partial_{(d)}\bigr)$ breaks into a direct sum of chain complexes indexed by the weight for the new grading.  Similarly, $\H_{0}\bigl(\MV_{\bullet}(X),\partial_{(d)}\bigr)$ is graded by polynomial degree.  By considering each graded piece individually, we can therefore choose a grading-preserving splitting $\varphi_{(d)} : \H_{0}\bigl(\MV_{\bullet}(X),\partial_{(d)}\bigr) \to \MV_{0}(X)$ of the projection $\tau_{(d)} : \MV_{0}(X) \to \H_{0}\bigl(\MV_{\bullet}(X),\partial_{(d)}\bigr)$, and for any such splitting we can iteratively choose a grading-preserving homotopy $\eta_{(d)} : \MV_{\bullet}(X) \to \MV_{\bullet+1}(X)$ satisfying $\varphi_{(d)}\tau_{(d)} = \id - [\partial_{(d)},\eta_{(d)}]$.

Let $\varphi' : \H_{0}\bigl(\MV_{\bullet}(X),\partial_{(d)}\bigr) \to \MV_{0}(X)$ be some other linear  splitting of the projection $\tau_{(d)}$ which does not necessarily preserve degree but does not increase it.  Then $\varphi'$ has a decomposition as $\varphi' = \sum_{w\geq 0} \varphi'_{w}$, where $\varphi'_{w}$ is homogeneous of weight $-w$.  The top part $\varphi'_{0}$ preserves degree, and necessarily also splits $\tau_{(d)}$.  Henceforth we identify $\varphi'_{0} = \varphi_{(d)}$, and choose a compatible grading-preserving homotopy $\eta_{(d)}$ as in the previous paragraph.  Set
$ \eta' = \eta_{(d)} + \eta_{(d)}\circ (\varphi' - \varphi) \circ \tau. $
By construction, $\eta' - \eta$ strictly lowers degree, and so $\eta'$ is degree-non-increasing.  A straightforward calculation implies that $\varphi'\circ \tau = \id - [\partial,\eta']$.

We now apply the Homological Perturbation Lemma~\ref{hpl} to the contraction
  $$ \begin{tikzpicture}[baseline=(L.base),anchor=base,->,auto,swap]
      \path node (L) {$\H_{0}\bigl(\MV_{\bullet}(X),\partial_{(d)}\bigr)$} ++(5,0) node (M) {$\bigl(\MV_{\bullet}(X),\partial_{(d)}\bigr)$} (M.mid) +(0,.075) coordinate (raise) +(0,-.075) coordinate (lower);
      \draw (L.east |- lower) to node {$\scriptstyle \varphi'$} (M.west |- lower);
      \draw (M.west |- raise) to node {$\scriptstyle \tau_{(d)}$} (L.east |- raise);
      \draw (M.south east) ..controls +(1,-.5) and +(1,.5) .. node {$\scriptstyle \eta'$} (M.north east);
    \end{tikzpicture}  $$
with deformation $\partial_{\cl} - \partial_{(d)}$, which is small because $\eta'$ is degree-non-increasing and $\partial_{\cl} - \partial_{(d)}$ strictly lowers degree.  By Remark~\ref{rmk.fact.hpl2}, we get a contraction of the form:
  $$ \begin{tikzpicture}[baseline=(L.base),anchor=base,->,auto,swap]
      \path node (L) {$\H_{0}\bigl(\MV_{\bullet}(X),\partial_{(d)}\bigr)$} ++(5,0) node (M) {$\bigl(\MV_{\bullet}(X),\partial_{\cl}\bigr)$} (M.mid) +(0,.075) coordinate (raise) +(0,-.075) coordinate (lower);
      \draw (L.east |- lower) to node {$\scriptstyle \varphi'$} (M.west |- lower);
      \draw (M.west |- raise) to node {$\scriptstyle \widetilde{\tau_{(d)}}$} (L.east |- raise);
      \draw (M.south east) ..controls +(1,-.5) and +(1,.5) .. node {$\scriptstyle \widetilde{\eta'}$} (M.north east);
    \end{tikzpicture}  $$
The formulas for $\widetilde{\tau_{(d)}}$ and $\widetilde{\eta'}$ guarantee that they are degree-non-increasing.  Let $\tau : \MV_{\bullet}(X) \to \H_{0}\bigl(\MV_{\bullet}(X),\partial_{\cl}\bigr)$ denote the natural projection.  Then $\tau\circ \varphi'$ is an isomorphism of filtered vector spaces $\Oo(\{\d s^{(d)}=0\}) = \H_{0}\bigl(\MV_{\bullet}(X),\partial_{(d)}\bigr) \cong \H_{0}\bigl(\MV_{\bullet}(X),\partial_{\cl}\bigr) = \Oo(\{\d s = 0\})$, and we have proven:
$$ \begin{tikzpicture}
  \path (0,0) node (Cx) {$\Oo(X)$} 
    (0,-2) node (cl) {$\Oo(\{\d s^{(d)} = 0\})$}
    (4,-2) node (q) {$\Oo(\{\d s = 0\})$};
  \draw[->] (Cx) -- node[auto,swap] {$\scriptstyle \tau_{(d)}$} (cl);
  \draw[->] (Cx) -- node[auto] {$\scriptstyle \tau$} (q);
  \draw[->%, blue
  ] (cl) .. controls +(.5,1) and +(.5,-1).. node[auto,swap] {$\scriptstyle \varphi'$} (Cx);
  \draw[%green!50!
  black] (cl) -- node[auto] {$\sim$} (q);
  \path
  (-9,-.5) node[anchor=west] {
  %\color{blue}
  For every filtered
  \tikz[baseline=(o)] \draw[%blue,
  ->] coordinate (o) ++(0,-3pt) coordinate (b) .. controls +(3pt,6pt) and +(3pt,-6pt).. ++(0,12pt);
   such that 
  $\tikz[baseline=(o)] \draw[->] coordinate (o) ++(0,-2pt)  .. controls +(3pt,6pt) and +(3pt,-6pt).. ++(0,11pt) -- ++(-2pt,0)  -- ++(0,-11pt); = \id$% 
   ,}
  ++(0,-18pt) node[anchor=west] {
  %\color{green!50!black}
  there exists a unique
  $\simeq$ such that
  $
   \tikz[baseline=(o)] \draw[->] coordinate (o) ++(0,-2pt) -- ++(-9pt,0) .. controls +(3pt,6pt) and +(3pt,-6pt).. ++(0,11pt) -- ++(3pt,0) -- ++(10pt,-11pt);   
   = \id
  $. 
  };
\end{tikzpicture}
$$

Put another way, we have given a construction turning any filtered splitting $\varphi'$ of $\tau_{(d)}$ into a filtered splitting $\varphi$ of $\tau$, and moreover shown that for any filtered splitting $\varphi$ of $\tau$ so constructed, there is a compatible filtered homotopy $\widetilde{\eta'}$.  
Moreover, by using $\widetilde{\eta'}$ as our homotopy, we can run the Homological Perturbation Lemma~\ref{hpl} in reverse to reconstruct $\varphi'$ from $\varphi$.  Thus the map $\bigl\{$filtered maps $\varphi':\Oo(\{\d s^{(d)}=0\})\to\Oo(X)$ splitting $\tau_{(d)}\bigr\} \to \bigl\{$filtered maps $\varphi:\Oo(\{\d s=0\})\to\Oo(X)$ splitting $\tau\bigr\}$ is an inclusion of finite-dimensional affine spaces.  Since we have constructed a filtered isomorphism $\Oo(\{\d s^{(d)}=0\}) \cong \Oo(\{\d s=0\})$ intertwining $\tau_{(d)}$ with $\tau$, the two spaces of possible splittings are isomorphic.  By finite-dimensionality, we conclude that every splitting $\varphi$ of $\tau$ comes from some splitting $\varphi'$ of $\tau_{(d)}$.

But our construction $\varphi' \mapsto \varphi$ produced a filtered homotopy $\eta = \widetilde{\eta'}$ compatible with $\varphi$.  Thus we can run the Homological Perturbation Lemma~\ref{hpl} again with deformation $\div$.  This produces the formula given in Theorem~\ref{thm.main} and completes the proof.

\subsection{Remark (dropping the condition that $s^{(d)}$ is nonsingular)}

The conclusion of Theorem~\ref{thm.main} ought to follow only from the condition that the critical locus $\{\d s = 0\}$ is zero-dimensional with no points at infinity.  If one adds the condition that the scheme $\{\d s = 0\}$ is reduced (i.e.\ that the Hessian of $s$ is nondegenerate at every critical point), then related results follow from the technique of Lefschetz thimbles.  We prefer our more algebraic approach, as it has a better chance of applying in infinite-dimensional situations, but without the nonsingularity of $s^{(d)}$ we are not aware of a general way to construct homotopies $\eta$ for which the deformation $\delta = \div$ is small.

\subsection{Proof of Theorem~\ref{thm.basis}} \label{proof.basis}

By Proof~\ref{proof.main}, it suffices to prove that $\H_{0}\bigl(\MV_{\bullet}(X),\partial_{(d)}\bigr)$ has the specified monomial basis.

A homogeneous degree-$d$ polynomial $s = s^{(d)}$ is \define{diagonal} if $s = \sum_{i} a_{i} \frac{(x_{i})^{d}}{d!}$.  It is clear that any homogeneous polynomial $s$ decomposes as $s = s_{\diag} + s_{\mix}$, where $s_{\diag}$ is diagonal and every monomial in $s_{\mix}$ contains at least two different variables.  We similarly decompose $\partial_{\cl} = \sum_{i} \frac{\partial s}{\partial x_{i}}\frac{\partial}{\partial \xi_{i}}$ as $\partial_{\cl} = \partial_{\diag} + \partial_{\mix}$.  Our strategy will be to consider $\partial_{\mix}$ as a (hopefully small) perturbation to $\partial_{\diag} = \sum_{i} a_{i} \frac{(x_{i})^{d-1}}{(d-1)!}\frac{\partial}{\partial \xi_{i}}$ in the Homological Perturbation Lemma~\ref{hpl}.

For $s$ generic, all $a_{i}$ are non-zero.  Then the complex $\bigl(\MV_{\bullet}(X),\partial_{\diag}\bigr)$ factors as a tensor product over $\CC$ of $2$-term complexes:
$$ \left( \CC[x_{1},\dots,x_{n},\xi_{1},\dots,\xi_{n}],  \sum_{i} a_{i} \frac{(x_{i})^{d-1}}{(d-1)!}\frac{\partial}{\partial \xi_{i}} \right) = \bigotimes_{i=1}^{N} \left( \CC[x_{i},\xi_{i}], a_{i} \frac{(x_{i})^{d-1}}{(d-1)!}\frac{\partial}{\partial \xi_{i}} \right) $$
Each tensorand has an obvious retraction onto its homology $H = \frac{\CC[x_{i}]}{(x_{i})^{d-1}}$.  Namely, we set $\varphi(x_{i}^{m}) = x_{i}^{m}$ for $m<d-1$, and choose the homotopy to be 
$$ \eta(x_{i}^{m}) = \begin{cases} 0, & m<d-1, \\ \frac{(d-1)!}{a_{i}} \,\xi_{i}\, x^{m-(d-1)}, & m \geq d-1. \end{cases} $$

We can tensor together the splittings to get a splitting $\varphi_{\diag}: H = \frac{\CC[x_{1},\dots,x_{n}]}{(x_{1}^{d-1},\dots,x_{n}^{d-1})} \to \CC[x_{1},\dots,x_{n}]$ of the projection $\tau_{\diag}$.  There is no functorial way to tensor together the homotopies of a retraction, but we can make an arbitrary choice for $\eta$, which by Remark~\ref{rmk.fact.hpl2} won't matter much.  A good choice for $\eta_{\diag} : V_{0} \to V_{1}$ is:
$$ \eta_{\diag}\bigl(x_{1}^{m_{1}}\cdots x_{n}^{m_{n}}\bigr) = 
\begin{cases} 0, & \text{all }m_{i} < d-1, \\
\displaystyle \frac{\sum_{i} \frac{\xi_{i}}{a_{i}} \bigl(\frac{\partial}{\partial x_{i}}\bigr)^{d-1}}{\sum_{i} \binom{m_{i}}{d-1}} \bigl(x_{1}^{m_{1}}\cdots x_{n}^{m_{n}} \bigr), & \text{some }m_{i} \geq d-1.
\end{cases}$$
In no formula will the choice of $\eta_{\diag}: V_{k} \to V_{k+1}$ for $k\geq 1$ appear, and we can always extend $\eta_{\diag}$ iteratively to the components of higher homological degree, while preserving the extra grading for which $\deg(x_{i}) = 1$ and $\deg(\xi_{i}) = d-1$.

We can now ask whether the perturbation $\delta = \partial_{\mix}$ is small with respect to the retraction $(V_{\bullet},\partial_{\diag},\frac{\CC[\vec x]}{(x_{i}^{d-1})},0,\varphi_{\diag},\tau_{\diag},\eta_{\diag})$; i.e.\ is the operator $(\id - \partial_{\mix}\eta_{\diag})$ invertible?  By construction, this operator preserves the grading, and so we decompose the retraction into a direct sum over weights $w\in \ZZ$, and ask for every $w$ whether $(\id - \partial_{\mix}\eta_{\diag})^{(w)}: V_{\bullet}^{(w)} \to V_{\bullet}^{(w)}$ is invertible, where $V_{\bullet}^{(w)}$ is the weight-$w$ piece of $\MV_{\bullet}(X)$. 

For each $w$, this latter question is about the invertibility of a finite-dimensional matrix.  Hence it is answered by whether the determinant of that matrix is or is not $0$, and this determinant is some polynomial in the coefficients of $s_{\mix}$.  On the other hand $(\id - \partial_{\mix}\eta_{\diag})^{(w)}$ is definitely invertible when $s_{\mix} = 0$.  Therefore, for each $w\in \ZZ$, $(\id - \partial_{\mix}\eta_{\diag})^{(w)}$ is invertible for generic~$s_{\mix}$.

At this point, we make an aside about vocabulary.  We have been using the word ``generic,'' which has a technical meaning in algebraic geometry.  A property holds \define{generically} if it holds on a dense Zariski-open subset, and \define{very generally} if it holds on a countable intersection of dense Zariski-open subsets.  Since we are working over an uncountable field, such an intersection is still uncountable and dense.  For example, we have shown that for very general $s_{\mix}$, and hence for a dense uncountable set, $\partial_{\mix}$ is a small deformation with respect to the retraction $(V_{\bullet},\partial_{\diag},\frac{\CC[\vec x]}{(x_{i}^{d-1})},0,\varphi_{\diag},\tau_{\diag},\eta_{\diag})$, as this smallness holds in the intersection of the countably many Zariski-open sets for which $(\id - \partial_{\mix}\eta_{\diag})^{(w)}$ is invertible.

On the other hand, for generic $s=s^{(d)}$ it follows from Remark~\ref{remark.serre} that  $\H_{\bullet}(V_{\bullet},\partial_{\diag} + \partial_{\mix})$ is $(d-1)^{n}$-dimensional and concentrated in degree $0$, and retains its $\ZZ$-grading by polynomial degree.  Let $M$ denote the highest degree of any non-zero class in $\frac{\CC[\vec x]}{(\partial_{i}s)}$.  Consider the deformation $\partial_{\mix}$ to the differential $\partial_{\diag}$ on the complex $\bigoplus_{w=0}^{M} V_{\bullet}^{(w)}$.  The smallness of $\partial_{\mix}$ on $\bigoplus_{w=0}^{M} V_{\bullet}^{(w)}$ requires just the invertibility of $M+1$ finite-dimensional matrices, and hence holds for generic $s$.  On the other hand, if $\partial_{\mix}$ is small, it follows from the Homological Perturbation Lemma~\ref{hpl} that $\H_{0}\bigl( \bigoplus_{w=0}^{M} V_{\bullet}^{(w)},\partial_{\diag} + \partial_{\mix}\bigr) \subseteq \H_{0}(V_{\bullet},\partial_{\cl}) = \frac{\CC[\vec x]}{(\partial_{i}s)}$ has a basis consisting of the representatives of the monomials $\{x_{1}^{m_{1}}\cdots x_{n}^{m_{n}}\}$ for which all $m_{i} < d-1$.  But  this basis, having the same size as the dimension of $\frac{\CC[\vec x]}{(\partial_{i}s)}$, must be a basis for the whole space.  This completes the proof.

\subsection{Corollary (explicit formulas from an \emph{ad hoc} choice of basis)} \label{my.thm4}

For very general homogeneous $s = \sum_{i_{1},\dots,i_{d} = 1}^{n} s_{i_{1}\dots i_{d}} \frac{x_{i_{1}}\cdots x_{i_{d}}}{d!}$, the projection $\tilde \tau :\Oo(X)  \to \H_{0}\bigl(\MV_{\bullet}(X),\partial_{\BV}\bigr) $ is given by:
  $$ \tilde\tau = \tau_{\diag} \bigl( {\id} - \partial_{\mix}\eta_{\diag} \bigr)^{-1} \sum_{\ell \geq 0} \left( \div\eta_{\diag} \bigl( \id - \partial_{\mix}\eta_{\diag} \bigr)^{-1}\right)^{\ell} $$
where:
\begin{align*}
   \partial_{\mix}\eta_{\diag}(x_{1}^{m_{1}}\cdots x_{n}^{m_{n}})  &
   = \begin{cases} 0, \hspace*{2.5in} & \text{  all }m_{i} < d-1, \hspace*{.5in} \\ 
   \displaystyle 
   \frac1{\sum_{i} \binom{m_{i}}{d-1}}
   \sum_{\substack{i_{1},\dots,i_{d-1},j \\ \text{not all equal}}} \frac{s_{i_{1}\dots i_{d-1}j}}{s_{j\cdots j}} \frac{x_{i_{1}}\cdots x_{i_{d-1}}}{(d-1)!}  \left(\frac{\partial}{\partial x_{j}}\right)^{d-1}(x_{1}^{m_{1}}\cdots x_{n}^{m_{n}}). \hspace*{-2in}
   \end{cases}
\\
 \div\eta_{\diag}\bigl(x_{1}^{m_{1}}\cdots x_{n}^{m_{n}}\bigr) & = \begin{cases} 0, \hspace*{2.5in} & \text{ all }m_{i} < d-1, \\
 \displaystyle \frac{1}{\sum_{i} \binom{m_{i}}{d-1}} \sum_{i} \frac{1}{s_{i\dots i}} \left(\frac{\partial}{\partial x_{i}}\right)^{d}\bigl(x_{1}^{m_{1}}\cdots x_{n}^{m_{n}} \bigr), & \text{ else.} \end{cases}
\end{align*}
Note that $\partial_{\mix}\eta_{\diag}$ preserves polynomial degree, and $\div\eta_{\diag}$ reduces it by $d$, so the sum over $\ell$ converges.  Similar but longer formulas apply when $s$ is allowed to be inhomogeneous.  If one is only interested in the values of $\tilde\tau$ on polynomials of fixed maximal degree, then the above formulas hold for generic $s$.

\subsection{Example (a case when Theorem~\ref{thm.basis} fails)}  The above formulas do not hold for all $s$.  For $s$ a generic quartic in two variables $x$ and $y$, Theorem~\ref{thm.basis} implies that $\Oo(\{ \d s = 0\})$ has as a basis the set of homology classes $\{[1],[x],[y],[x^{2}],[xy],[y^{2}],[x^{2}y],[xy^{2}],[x^{2}y^{2}]\}$.  But for $s(x,y) = x^{4} + 2x^{3}y + 2xy^{3} + y^{4}$, for example, $[x^{2}y^{2}] = [\frac1{12} \bigl( (2y - x)\partial_{x}s + (2x - y) \partial_{y}s\bigr)] = 0$ in $\Oo(\{\d s = 0\})$ and thus cannot be a basis element.  This example illustrates that the genericity assumption on $s$ cannot be dropped from Theorem~\ref{thm.basis}.

\subsection{Remark (as far as algebra can go?)}

Since the integration pairing $\H_{n}\bigl(\CC^{n};\{\Re(s)\ll 0\}\bigr) \otimes \H_{0}\bigl(\MV_{\bullet}(X),\partial_{\BV}\bigr) \to \CC$ between contours and observables is perfect~\cite{MR2443034}, for a general contour $\gamma$ and fixed action $s$ Theorems~\ref{thm.main} and~\ref{thm.basis} and Corollary~\ref{my.thm4} are as much as pure algebra can say about the values of integrals.  That said, in special cases there is often more that can be computed by studying the symmetry of the problem.  Moreover, for fixed $f\in \Oo(X)$ and contour $\gamma$, one can write differential equations describing how $I_{s}(f) = \int_{\gamma}f\,e^{s}\,\dVol$ varies as a function of the coefficients of $s$.  (If $s$ is changed by a small amount, $\gamma$ remains allowable.)  For instance, there is a one-parameter family interpolating between $s$ and $s^{(d)}$, and (provided $s^{(d)}$ is nonsingular) Proof~\ref{proof.main} identifies the quantum BV homologies for all members of this family with $\Oo(\{\d s^{(d)}=0\}) \cong\CC^{(d-1)^{n}}$; different members of the one-parameter family give different integration maps out of $\Oo(\{\d s^{(d)}=0\})$, which are related by an explicit differential equation.

A related question is to understand the values of $I_{s}(1) = \int_{\gamma}e^{s}\dVol$.  For fixed $\gamma$, $I_{s}(1)$ is an $\mathfrak{sl}(n,\CC)$-invariant of $s$, transforming in a specific weight space for the center of $\mathfrak{gl}(n,\CC)$.  Indeed, $I_{s}(1)$ solves a differential equation  making it a branch of a hypergeometric function of the polynomial invariants of $s$.  This approach has been pursued in~\cite{MR2593033,MR2730148}.

\section{So, can we compute nonperturbative path integrals?} \label{section4}

In this section we address to what extent the techniques we have developed so far apply to the infinite-dimensional integrals that appear in the path-integral approach to quantum field theory.  The Feynman-diagrammatics described in Example~\ref{eg.Feynman} have proven immensely useful in high-energy physics and mathematics, so we will focus on the question of translating into infinite dimensions our approach to nonperturbative integrals.
We will not prove any results, but simply outline some directions for further research.

We should not expect there to be enough patterns in the algebraization of finite-dimensional integral problems to be able to take the limit as $n\to \infty$ at the end of the problem.  Rather, we can hope to begin with a complex playing the role of $\BV_{\bullet}(X) = \bigl(\MV_{\bullet}(X),\partial_{\BV}\bigr)$ and study it with the Homological Perturbation Lemma~\ref{hpl}.  Doing so would give the algebraic part of the integral; then we could \emph{define} an allowable contour for the infinite-dimensional integral problem as a map $\H_{0}\bigl(\MV_{\bullet}(X),\partial_{\BV}\bigr) \to \CC$.

\subsection{Remark (choices and ultraviolet divergences)}

The new feature in infinite-dimensional problems is that one must make choices where none were necessary in finite dimensions.  Rather than giving a general story, we focus on a simplified picture, in which the space replacing $X = \CC^{n}$ to be integrated over is an infinite-dimensional vector space  of sections of some vector bundle.    The first step in generalizing the construction above is to come up with a reasonable version of the graded algebra $\MV_{\bullet}(X)$.  Recall from Example~\ref{eg.smooth} that when $X$ is finite-dimensional, $\MV_{\bullet}(X) = \Gamma\bigl( \T^{\wedge\bullet}X\bigr)$, and for $X = \AA^{n}$ we took polynomial sections in Example~\ref{eg.main}.  For finite-dimensional vector spaces $X$, we thus have $\MV_{\bullet}(X) = \Lambda^{\bullet}(X) \otimes \Sym(X^{*})$, where $X^{*}$ is the dual vector space to $X$.  When $X$ is infinite-dimensional, we can try to take this as a definition, but now two choices must be made: first, which dual space to take, and second, how to complete the myriad tensor products present in the symmetric and antisymmetric powers.

Unfortunately, it is rare to find such choices so that both ``$\partial_{\cl}$'' and ``$\div$'' are defined on $\MV_{\bullet}(X)$.  In general, to define $\partial_{\cl}$ requires that the tensor products be completed appropriately.  On the other hand, the invariant definition of $\div$ is as an extension of the map that pairs $X$ with $X^{*}$, and this pairing is generally defined on the algebraic tensor product $X\otimes X^{*}$ but not on whatever completion is required to define $\partial_{\cl}$.

This problem arises when defining perturbative integrals as well, and in that context it is called the problem of \define{ultraviolet divergences}.  In the perturbative context the solution is reasonably understood, and goes by the name \define{renormalization theory}.  The idea is to define $\div$ on the algebraic tensor product and then somewhat arbitrarily choose an extension of it to whatever tensor completion is necessary.  Almost certainly, such $\div$ will not commute with $\partial_{\cl}$, and so the naive guess for $\partial_{\BV}$ will not be a differential.  But in the perturbative integral, the hoped-for differential is (after rescaling by $\hbar$) $\partial_{\BV} = \partial_{\cl} + \hbar \div$, and so the failure to square to zero is order $\hbar$.  One then modifies $\partial_{\cl}$ by a term which is order $\hbar$ in such a way that $\partial_{\cl}$ no longer squares to $0$, but so that
 $\partial_{\BV}^{2} = O(\hbar^{2})$.  After another modification, $\partial_{\BV}^{2} = O(\hbar^{3})$.  In good situations, one can repeat this process infinitely, so that for a modified $\partial_{\cl}$ (or, what is equivalent, a modified action $s$) one can define a differential $\partial_{\BV}$.  Once defined, one can use the Homological Perturbation Lemma~\ref{hpl} to study $\partial_{\BV}$ in terms of the unmodified $\partial_{\cl}$, and the answer is given by Feynman diagrams.  This understanding of renormalization theory underpins \cite{MR2778558,costellogwilliam}, for example. \label{choices}
 
\subsection{Example (Chern--Simons Theory)}

The presence of ultraviolet divergences is a major obstruction to transporting the homological understanding of integration to the infinite-dimensional setting.  But it seems to be the only one.  Provided that one  works with an algebra $\MV_{\bullet}(X)$ that deserves to be thought of as an algebra of ``polynomial multivector fields,'' one should still have the local nilpotence necessary in Theorems~\ref{thm.main} and~\ref{thm.basis} to reduce the ``quantum'' problem of understanding the homology of $\partial_{\BV}$ to the ``classical'' problem of understanding the homology of $\partial_{\cl}$.  As an illustration, we discuss the well-studied example of quantum Chern--Simons Theory, the path integral for which conjecturally computes Reshetikhin--Turaev invariants of knots and three-manifolds \cite{MR990772,MR1036112,KolyaShort}.

Again we present only a very simplified version.  Focusing on knot invariants, we fix our spacetime manifold to be the three-sphere $M = S^{3}$, and we choose a compact Lie group $G$ with Lie algebra $\mathfrak g$.  Then the naive space of fields to integrate over is $\Gamma = \Omega^{1}(M) \otimes \mathfrak g$.  The \define{Chern--Simons action functional} $s$ has as its critical locus the \define{flat} sections $\gamma \in \Gamma$, i.e.\ those satisfying the Maurer--Cartan equation $\d \gamma(w_{1},w_{2}) = \frac12 [\gamma(w_{1}),\gamma(w_{2})]$ for all $x\in M$ and $w_{1},w_{2} \in \T_{x}M$.  There is a group $\Ggg = \hom(M,G)$ acting nonlinearly on $\Gamma$.  It does not preserve $s$, but it does preserve the one-form $\d s$, which is to say the $\Ggg$ preserves the differential $\partial_{\cl} = $ ``contract with $\d s$.''  Thus $\d s$ makes sense as a closed (but not exact) one-form on $\Gamma / \Ggg$, and with the correct normalization it has integer periods.  It is really over $\Gamma / \Ggg$ that  Chern--Simons Theory integrates.

Homological algebra is well-adapted to make sense of quotient spaces.  Just like the classical BV complex $\bigl(\MV_{\bullet}(X),\partial_{\cl}\bigr)$ is a ``derived intersection'' computing certain Tor groups, \define{derived quotients} can be defined as the chain complexes computing certain Ext groups.  In infinite dimensions to do this technically requires much work, mostly of the ``making choices'' form discussed above.  Then derived quotients can be combined with the formation of odd cotangent bundles in a certain homological version of Marsden--Weinstein reduction.

When $\Ggg$ is replaced by the Lie algebra $\Omega^{0}(M) \otimes \mathfrak g$ (thought of as an infinitesimal group), the result of these derived operations is reasonably well-known.  The answer is that
$$\MV_{\bullet}\Bigl(\Gamma/(\Omega^{0}(M) \otimes \mathfrak g)\Bigr) =  \Sym\Bigl( \bigl(\Omega^{0}(M)[0] \oplus \Omega^{1}(M)[1] \oplus \Omega^{2}(M)[2] \oplus \Omega^{3}(M)[3]\bigr)\otimes \mathfrak g[-1]\Bigr) $$
where the numbers in brackets denote shifts in homological grading, and  the symmetric algebra construction is interpreted in the graded sense.  
In writing this, we have made specific choices of the type described in Remark~\ref{choices}: we used the orientation on $M$ to identify $\bigl(\Omega^{k}(M)\bigr)^{*} = \Omega^{3-k}(M)$ and a choice of Killing form on $\mathfrak g$ to identity $\mathfrak g^{*} = \mathfrak g$.   The differential $\partial_{\cl}$ combines a ``de Rham'' part and a ``Chevalley--Eilenberg'' part.  A more natural origin of this infinite-dimensional derived space is described in \cite{MR1432574}.

To be able to apply a version of Theorem~\ref{thm.main} from this paper, it will probably be necessary to use not the infinitesimal group $\Omega^{0}(M) \otimes \mathfrak g$ but the full gauge group $\Ggg$, just as Theorem~\ref{thm.main} uses the full scheme $\AA^n$ and not the infinitesimal neighborhood of the critical locus $\{\d s = 0\}$.  Making such a change will surely require the derived geometry of~\cite{MR3090262}.

\subsection{Remark (Volume Conjecture and analytic continuation)}

One final remark is in order.  There has been continuing interest in the analytic continuations of Chern--Simons theory to non-compact gauge groups \cite{MR2809462}.  One motivation comes from the ``Volume Conjecture'' \cite{MR1434238}: in Witten's original path-integral description of the Jones polynomial \cite{MR990772}, one imagines integrating over connections valued in $\mathrm{SU}(2)$, but the volume conjecture predicts that in a certain limit the Jones polynomial is dominated by certain connections valued instead in $\mathrm{SL}(2,\RR)$.

The method of steepest descent implies that many integrals of the form $\int f e^{s}$ can be dominated by imaginary critical points of $s$, and Theorem~\ref{thm.main} says that this is a purely algebraic result, and so should apply even to as-yet-to-be-defined infinite-dimensional integrals.  (If $s$ has degenerate critical points, the usual method of steepest descent does not give an answer, but Theorem~\ref{thm.main} still allows one to work with the non-reduced scheme-theoretic critical locus.)  
For Theorem~\ref{thm.main} to apply, our observables must be polynomial: applying the same techniques but with $\mathscr{C}^{\infty}$ observables would give dramatically different (and much less algebraic) answers.  

In some versions of Chern--Simons theory with non-compact gauge group, one works with smooth, rather than polynomial, functions.
 To a polynomial, $\mathrm{SU}(2)$ and $\mathrm{SL}(2,\RR)$ are essentially indistinguishable, but to a smooth function they are very different.  Sometimes (\cite{MR2809462}, for example) it is useful to play algebraic and smooth versions of a group off each other.  But it is also important to keep them separate.  In particular, it is reasonable to predict that there is a purely algebraic version of Chern--Simons theory in which the groups appearing are always treated as algebraic objects, and an algebraic, nonperturbative integral over $\mathrm{SU}(2)$-connections will, according to the evidence in this paper, be controlled by the scheme of flat $\mathrm{SL}(2, \CC)$-connections, and not just its real points.

%\bibliography{ReferencesWithLinks}{}
%\bibliographystyle{alpha}
%\addcontentsline{toc}{section}{References}

\end{document}